%% file: paper.tex
\documentclass[useAMS,usenatbib]{mnras}
\usepackage{graphicx,natbib,amssymb,amsmath,stmaryrd,makecell}
\usepackage{txfonts}
\usepackage{xcolor}
\usepackage{hyperref}
\usepackage{xspace,dsfont}
\usepackage{caption, lipsum, comment}
\usepackage{bm}
\usepackage{orcidlink}

\graphicspath{{Figures/}}

\input{Befehle}

\newcommand{\oDnu}{{\mathcal{\hat{D}}_{\nu}}}

\newcommand{\oOnu}{{\mathcal{\hat{O}}_{\nu}}}

\newcommand{\oOx}[1]{{x^{#1}\partial_x^{#1}}}

\renewcommand{\Ne}{N_{\rm e}}

\usepackage{comment}

\newcommand{\Bb}[4]{{^{#1}_{#2}}\mathcal{\hat{B}}^{#3}_{#4}}
\newcommand{\Bbm}[3]{{^{#1}_{#2}}\mathcal{\hat{B}}^m_{#3}}
\newcommand{\Bbom}[2]{{^{0}_{#1}}\mathcal{\hat{B}}^m_{#2}}

\newcommand{\Bbob}[3]{{^{-1}_{#1}}\mathcal{\hat{B}}^{#3}_{#2}}

\newcommand{\Bbomb}[2]{\Bbob{#1}{#2}{m}}

\newcommand{\Kk}[4]{{^{#1}_{#2}}\mathcal{K}^{#3}_{#4}}

\newcommand{\Dbo}[3]{{{_{#1}}\mathcal{\hat{D}}^{#3}_{#2}}}

\newcommand{\Dbom}[2]{\Dbo{#1}{#2}{m}}

\newcommand{\Yy}[2]{{_{#1}}Y_{#2}}

\newcommand{\bra}[1]{\langle #1 \vert}
\newcommand{\ket}[1]{\vert #1 \rangle}

\usepackage{bbm}

\title[Boost operator approach to rpSZ]
{Boost operator approach to the relativistic polarized SZ effect}

\author[]
{
Erik~Rosenberg\,\orcidlink{0000-0003-3484-5645}$^{1}$\thanks{E-mail:Erik.Rosenberg@Manchester.ac.uk}
and
Jens~Chluba\,\orcidlink{0000-0003-3725-6096}$^{1}$
\\
$^{1}$Jodrell Bank Centre for Astrophysics, Alan Turing Building, University of Manchester, Manchester M13 9PL
}

\begin{document}

\date{Accepted 2025 --. Received 2025}

\maketitle

\begin{abstract}
We extend the recent boost operator formalism for relativistic Compton scattering calculations to also account for polarization. This allows us to provide general, exact expressions for the polarized Sunyaev-Zeldovich (SZ) effect sourced both kinematically and from intrinsic anisotropies of the Cosmic Microwave Background (CMB). The results are given in terms of rational operator functions that can be used to generate distortion spectra that describe the general SZ signal, reproducing the classical polarized SZ results in the appropriate limits. Our derivation allows for clear separation of physical effects in the generation of polarized SZ, and beyond the SZ application provides a general description of the Compton collision term in the Doppler-dominated regime. Through direct computation of important example cases we further illustrate the power of this new method.
\end{abstract}
%------------------------------------------------------------


%---------------------------
\section{Introduction}
\label{sec:Intro}
%---------------------------
The problem of radiative transport, describing the evolution of a radiation field in the presence of a population of moving scattering particles, is fundamental in astrophysics and cosmology.
For Compton scattering, one of the major applications is  the Sunyaev-Zeldovich (SZ) effect \citep{Zeldovich1969, Sunyaev1980}, which describes the scattering of anisotropies of the cosmic microwave background (CMB) off of free electrons in galaxy clusters, and has developed into an important probe of the large-scale structure \citep[e.g.][]{Carlstrom2002, Mroczkowski2019}.
When the scattering population is relativistic, as can be the case with electron scattering in high-temperature clusters, these calculations are complicated by the need to include relativistic effects, generally treated by asymptotic expansion in the cluster velocity and electron temperature \citep[e.g.,][]{Wright1979, Rephaeli1995, Itoh98, Sazonov1998}. 
Accounting for the generation of polarization \citep[e.g.][]{Challinor2000,
Itoh2000Pol} can be done with the same formalism but with further complication.

In the standard calculation, many integrals over the scattering angles of the electrons and photons have to be carried out, making the derivation quite demanding. 
An alternative framework for these calculations, termed the boost operator approach, has recently been set forth in \citet{ChlubaBO25} and applied to the (intensity) SZ effect in \citet{ChlubaBO25b}.
The boost operator approach is rooted in the exact treatment of light aberration -- accounting for the (relativistic) effect of our proper motion with respect to the CMB frame on the CMB anisotropies.
The aberration effect can be computed using the aberration kernel \citep[e.g.][]{Challinor2002, Chluba2011ab, Dai2014}, which describes how CMB temperature anisotropies or other (frequency-independent) fields on the sphere transform under relativistic boosts.
This idea was generalized by \citet{ChlubaBO25} to define the boost operator, which also accounts for relativistic Doppler boosting and can therefore be applied to understand the transformation of frequency-dependent fields.

In this framework, quantities of the photon field (e.g., occupation number, intensity, etc) are transformed using the boost operator into the electron rest frame, where the effects of scattering can be computed significantly more simply, and afterwards transformed back into the lab frame to obtain the final result.
This approach has allowed an elegant derivation of the relativistic SZ effect, with exact expressions in terms of products of boost operators, which themselves are rational operator functions that can be generated using recursion relations \citep{ChlubaBO25}. Furthermore, when rewriting the operators in powers of the electron momenta $p$, one can generate all higher-order expressions as powers of the energy shift operator, $\oOnu=-\nu\partial_\nu$, eliminating the need to perform computationally intensive integrals \citep{ChlubaBO25b}.

In this work, we  extend the boost operator procedure into polarization. The main additional complication in treating polarization is the need to track the polarized states and their mixing in scattering events. In the electron rest frame this can be easily achieved by following the method of \citet{Hu1997}, as long as recoil terms ($\simeq h\nu/\me c^2$) can be neglected. This is akin to treating the Compton scattering problem in the Doppler-dominated limit of the Compton scattering kernel \citep{CSpack2019}.
However, this means that we also have to properly treat the transformation of spin-$\pm2$ fields with the boost operator. 
To properly keep track of the polarization states, we will also need to perform an additional rotation of the result into a general frame, a step that was not necessary in the work of \citet{ChlubaBO25b}.
Altogether this allows us to derive the relativistic polarized SZ effect, reproducing previous results \citep[e.g.,][]{Itoh2000Pol, Challinor2000} and extending them to any temperature or polarization anisotropy for both isotropic electron distributions and isotropic distributions with additional bulk motion. The latter leads to kinematic corrections to the scattering process, required to obtain the classical pSZ result \citep{Sunyaev1980}.

In Sec.~\ref{sec:psz} we derive the scattering operator for a single scattering event with general electron direction, followed by the general direction-averaged operators for scattering off of an isotropic electron distribution and a thermal election cloud with a peculiar velocity.
In Sec.~\ref{sec:cmb_apps} we explicitly compute the example cases of scattering off the CMB monopole, quadrupole and octupole.
The appendices contain useful derivations relating to the spin-weighted spherical harmonics and Wigner-D matrices as well as recursion relations for the computation of the relevant aberration kernels.

\vspace{-4mm}
\section{Scattering of polarized radiation}
\label{sec:psz}
%---------------------------
To calculate the polarized SZ effect we effectively repeat the procedure of \citet{ChlubaBO25b} while keeping track of polarized states. The intensity matrix defines the usual Stokes parameters $I$, $Q$, and $U$ (we neglect circular polarization, which is not generated by Thomson scattering) as
%---------------------------
\begin{equation}
    \mathcal{I} = 
    \begin{pmatrix}
    I+Q & U \\
    U & I-Q
    \end{pmatrix}
\end{equation}
%---------------------------
using an $\bm{\hat e}_k=\{\bm{\hat e}_\theta, \bm{\hat e}_\varphi\}$ basis for radial radiation. Because $Q$ and $U$ defined as such mix under rotations, we instead work with $Q \pm iU$ which transform as spin-$\pm2$ fields. Our polarized state vector is then 
%---------------------------
\begin{equation}
    \bm{n}(\nu, \vgh) = \begin{pmatrix} n^I & n^Q+in^U & n^Q-in^U \end{pmatrix}^{\rm T}
\end{equation}
%---------------------------
where $\nu$ and $\vgh$ denote photon frequency and direction while $\bm{n}$ indicates that we choose to work with the occupation numbers corresponding to the usual Stokes intensities (e.g., $I\equiv [2h\nu^3/c^2]\,n$). For convenience we will refer to the polarization components as $n^\pm \equiv n^Q \pm in^U$ given they are spin-weight $s=\pm2$ quantities. 

\vspace{-3mm}
\subsection{Single Scattering}
\label{sec:single-scattering}
The Thomson collision term for polarized radiation in the rest-frame of a scattering electron is given by \citet{Hu1997}:
%---------------------------
\begin{align}
    \left.\frac{\id n'^i(\nu', \vgh')}{\id\tau'}\right|_{\mathrm{T}} &=
    \delta^{i0}\,{n'}^i_0(\nu')-{n'}^i(\nu',\vgh')
\nonumber\\
    &\qquad + \int \frac{\id\Omega'}{10}\sum_{j m} P^{ij}_{(m)}(\vgh', \Omega') \, n'^j(\nu', \Omega') 
    \label{eq:thomson}
\end{align}
%---------------------------
Here $i$ denotes the index of the polarization vector, $P$ is the mixing matrix describing the scattering that is given by \citet{Hu1997}, and the sum over $m$ runs from $-2$ to $2$, since Thomson scattering couples to the quadrupole anisotropy of the photon field. Primes are used to denote that these are all electron rest frame quantities. We also note that ${n'}^0_0(\nu')$ is the intensity monopole, ${n'}^0_0=\int n'_I \id \vgh'/[4\pi]\equiv {n'}^0_{00}/\sqrt{4\pi}$. 
The $P$-term, describing quadrupole scattering into the line of sight, is the only change with respect to the unpolarized case. 
The $P$ matrix can be written 
%---------------------------
\begin{equation}
    P^{ij}_{(m)}(\vgh', \Omega') = c^{ij}\, \Yy{s^i}{2m}(\vgh')\, \Yy{s^j}{2m}^*(\Omega')
\end{equation}
%---------------------------
for constants $c^{ij}$. Here, $\Yy{s^i}{\ell m}$ are the spherical harmonics of spin weight $s_i = \{0, +2, -2\}$. The constants $c^{ij}$, given in \citet{Hu1997}, can be written as
%---------------------------
\begin{equation}
    \bm{c} = 
    \begin{pmatrix}
    1 &-\frac{\sqrt{6}}{2} &-\frac{\sqrt{6}}{2}
    \\
    -\sqrt{6} & 3 & 3
    \\
    -\sqrt{6} & 3 & 3
    \end{pmatrix}
%    \begin{pmatrix} 1 & -\sqrt{6} & -\sqrt{6}  \end{pmatrix}
%    \otimes
%    \begin{pmatrix} 1 & -\sqrt{\frac{3}{2}} & -\sqrt{\frac{3}{2}} \end{pmatrix}.
\end{equation}
%---------------------------
In the last term of Eq.~\eqref{eq:thomson}, we insert $n'^j(\Omega') = \sum_{\ell m} {_{s^j}}Y_{\ell m}(\Omega') \,{n'}^j_{\ell m}$ and then evaluate the integral using orthonormality of spin-weighted spherical harmonics. This  lets us write the last term of Eq.~\eqref{eq:thomson} as 
%---------------------------
\begin{equation}
    \left.\frac{\id n'^i(\nu', \vgh')}{\id\tau'}\right|_{\mathrm{T, in}} = \sum_{j m} \frac{c^{ij}}{10}\, {_{s^i}}Y_{2m}(\vgh')\, {n'}_{2m}^{j}.
    \label{eq:inscatter}
\end{equation}
%---------------------------
Using the boost operator, $\Bbm{d}{s}{\ell\ell'}(\nu, \beta)$, we now want to rewrite the rest-frame scattering $\id n'^i/\id \tau'$ in terms of the CMB-frame quantities $n^i$.
As defined in \citet{ChlubaBO25}, the boost operator connects the spin harmonic coefficients in one frame to another as\footnote{The `$-s$' in the boost operator stems from the definition of the aberration kernel ${_s^d}\mathcal{K}^{mm'}_{\ell \ell'}$ \citep{Dai2014} when $X$ has spin-weight $s$.}
%---------------------------
\bealf{
{X'}^{i}_{\ell m}(\nu) &= \sum_{\ell'} \Bbm{d_i}{-s_i}{\ell\ell'}(\nu, \beta) \,X^{i}_{\ell'm}(\nu)
}
%---------------------------
for a quantity $X^i$ of spin-weight $s_i$ and Doppler weight $d_i$.
Here, the speed is given by $\beta=\varv/c$ and it is assumed that the motion is along the $z$-axis.
Noting that occupation number has Doppler weight $d=0$ and using the simplification of Eq.~\eqref{eq:inscatter} we rewrite Eq.~\eqref{eq:thomson} as
%---------------------------
\bealf{
&\left.\frac{\id  n'^i(\nu',\vgh')}{ \id \tau'}\right|_{\rm T}
=\left(\sum_{\ell'} {_0}Y_{00}(\vgh')\,\Bb{0}{0}{0}{0\ell'}(\nu',\beta) \,n^0_{\ell' 0}(\nu')\right)\delta^{i0}
\nonumber\\
& \qquad\qquad + \sum_{ m \ell' j} \frac{c^{ij}}{10}\, {_{s^i}}Y_{2m}(\vgh')\, \Bbom{-s^j}{2\ell'}(\nu', \beta)\,n_{\ell' m}^j(\nu')
\nonumber\\
&\qquad\qquad\qquad - \sum_{\ell m \ell'} {_{s^i}}Y_{\ell m}(\vgh') \,\Bbom{-s^i}{\ell\ell'}(\nu', \beta)\, n^i_{\ell'm}(\nu')
\\ \nonumber 
&=\sum_{\ell m \ell' j}\!\left\{
(\delta^{j0}\delta_{\ell0}-1)\,\delta^{ij}+\frac{c^{ij}\delta_{\ell2}}{10}\right\} {_{s^i}}Y_{\ell m}(\vgh')\, \Bbom{-s^j}{\ell\ell'}(\nu', \beta) \,n_{\ell' m}^j(\nu')
}
%---------------------------
This is the same as the expression given in \citet{ChlubaBO25} when neglecting polarization, i.e., $c^{ij}=\delta_{i0}\delta_{j0}$. The sum over polarization states $j\neq 0$ in the term for photons scattering into the line of sight and accompanying coefficients $c^{ij}$ are the main generalisation with the appropriate spin-weights in the boost operators.

All that remains now is to transform this expression back into the lab frame. Since $\id  n'^i(\nu',\vgh', \vb)/\id\tau'$ has Doppler-weight $d=-1$, we simply have to replace $${_{s^i}}Y_{\ell' m}(\vgh')\rightarrow {_{s^i}}Y_{\ell m}(\vgh)
\,\Bbomb{-s^i}{\ell\ell'}(\nu, -\beta)/\gamma$$
and sum over $\ell$. The factor of $1/\gamma$ arises from the transformation of the optical depths \citep{ChlubaBO25}, where $\gamma=\sqrt{1+p^2}$ is the Lorentz factor for $\beta=p/\gamma$. Introducing the {\it Doppler operator}
%---------------------------
\bealf{
\Dbom{ij}{\ell \ell' \ell''}(\nu, \beta)
&=
\frac{\Bbomb{-s^i}{\ell\ell'}(\nu, -\beta)\,\Bbom{-s^j}{\ell'\ell''}(\nu, \beta)}{\gamma},
}
%---------------------------
then yields
%---------------------------
\bealf{
\nonumber
&\left.\frac{\id  n^i(\nu,\vgh)}{ \id \tau}\right|_{\rm T}
=\sum_{\ell m\ell'\ell'' j}
\!\!{_{s^i}}Y_{\ell m}(\vgh)
\left\{
(\delta^{j0}\delta_{\ell'0}-1)\,\delta^{ij}+\frac{c^{ij}\delta_{\ell'2}}{10}\right\} 
\Dbom{i j}{\ell \ell' \ell''} n_{\ell'' m}^j
}
%---------------------------
as the general result for the effect of a single boost to the scattering of photons by moving electrons.

The expression can be further simplified by using identities for the boost operator. With the Doppler weight raising operation, one can write \citep{ChlubaBO25}
%---------------------------
\bealf{
\frac{\Bbomb{s}{\ell\ell'}(\nu, -\beta)}{\gamma} &=
\Bbom{s}{\ell\ell'}(\nu, -\beta)
-\beta \bigg[
{_{s}}C^m_{\ell+1}\,
\Bbom{s}{\ell+1\ell'}(\nu, -\beta)
\\ \nonumber 
&\qquad
+\frac{sm}{\ell(\ell+1)}\,\Bbom{s}{\ell\ell'}(\nu, -\beta)
+{_s}C^m_{\ell}\,\Bbom{s}{\ell-1\ell'}(\nu, -\beta)
\bigg]
}
%---------------------------
with $\ell \,{_s}C_\ell^m=\sqrt{(\ell^2-m^2)(\ell^2-s^2)/(4\ell^2-1)}$. This then implies
%---------------------------
\bealf{
\nonumber
\sum_{\ell'}\Dbom{i i}{\ell \ell' \ell''} &=
\delta_{\ell\ell''}
-\beta \bigg[
{_{s_i}}C^m_{\ell+1}\,\delta_{\ell+1,\ell''}
-\frac{s_i \,m}{\ell(\ell+1)}\,\delta_{\ell\ell''}
+{_{s_i}}C^m_{\ell}\,\delta_{\ell-1,\ell''}
\bigg],
}
%---------------------------
which avoids having to take an infinite sum over intermediate states $\ell'$.
This then yields the compact expression
%---------------------------
\bsub
\label{eq:final_general_expressions}
\bealf{
\left.\frac{\id  n^i(\nu,\vgh)}{ \id \tau}\right|_{\rm T}
&=\sum_{\ell m \ell'' j}
\!\!{_{s^i}}Y_{\ell m}(\vgh)
\,{_{ij}}\hat{\mathcal{S}}^{m}_{\ell \ell''}(\nu, \beta)\,n_{\ell'' m}^j(\nu)
\\
{_{ij}}\hat{\mathcal{S}}^{m}_{\ell \ell''}(\nu, \beta)
&=\Dbo{0 0}{\ell 0 \ell''}{m}\,
\delta^{i0}\delta^{j0}
+\frac{c^{ij}}{10}\,\Dbom{i j}{\ell 2 \ell''}-\delta_{\ell\ell''}\delta^{ij}
\\ \nonumber
&\qquad 
+\beta \bigg[
{_{s_i}}C^m_{\ell+1}\,\delta_{\ell+1,\ell''}
-\frac{s_i \,m}{\ell(\ell+1)}\,\delta_{\ell\ell''}
+{_{s_i}}C^m_{\ell}\,\delta_{\ell-1,\ell''}
\bigg]\delta^{ij}.
}
\esub
%---------------------------
Here we defined the Thomson scattering operator, ${_{ij}}\hat{\mathcal{S}}^{m}_{\ell \ell''}(\nu, \beta)$, for a single boost along the $z$-axis including all orders in $\beta$.

\vspace{-3mm}
\subsection{General boost direction}
%---------------------------
To allow for general boost direction, we need to rotate the spherical harmonic coefficients from a general frame into the frame where $\vb$ is aligned with the $z$-axis. This can be achieved using the Wigner D-matrices (Appendix~\ref{app:sYlm_explicit} and \ref{app:trans_law_Ylm_alm}). From Eq.~\eqref{eq:trans_alm}, we have
%---------------------------
\bealf{
n'_{\ell m}=\sum_{m'} D^{\ell *}_{m' m}(\vbh)\,n_{\ell m'}
}
%---------------------------
where $D^\ell_{m' m}(\vbh) = D^\ell_{m' m}(\phi, \theta, \chi)$ is the Wigner D-matrix representing the rotation of the coordinate system to the system where the $z$-axis is aligned with $\vbh = (\theta, \phi)$ in the lab system $S$. The additional roll angle $\chi$ can be freely chosen. Once in the special frame $S'$, we can use the expression Eq.~\eqref{eq:final_general_expressions} to obtain the scattered radiation
%---------------------------
\bealf{
\label{eq:final_general_expressions_Sp}
\left.\frac{\id  n^i(\nu,\vgh')}{ \id \tau}\right|^{S'}_{\rm T}
&=\sum_{\ell m \ell'' m' j}
\!\!{_{s^i}}Y'_{\ell m}(\vgh')
\,{_{ij}}\hat{\mathcal{S}}^{m}_{\ell \ell''}(\nu, \beta)\,D^{\ell'' *}_{m'' m}(\vbh)\, n_{\ell'' m''}^j(\nu),
}
%---------------------------
where here $\vgh'$ is the required direction in $S'$. Using Eq.~\eqref{eq:trans_Ylm}, we can then also write
%---------------------------
\bsub
\bealf{
\label{eq:final_general_expressions_S}
\sum_{m}
&{_{s^i}}Y'_{\ell m}(\vgh')
\,{_{ij}}\hat{\mathcal{S}}^{m}_{\ell \ell''}(\nu, \beta)\,D^{\ell'' *}_{m'' m}(\vbh)
=
\sum_{m}
{_{s^i}}Y_{\ell m}(\vgh)
\,{_{ij}}\hat{\mathcal{S}}^{m m''}_{\ell \ell''}(\nu, \vb)
\\[1mm]
&{_{ij}}\hat{\mathcal{S}}^{m m''}_{\ell \ell''}(\nu, \vb)=
\sum_{m'}\,D^{\ell}_{m m'}(\vbh)
\,{_{ij}}\hat{\mathcal{S}}^{m'}_{\ell \ell''}(\nu, \beta)\,D^{\ell'' *}_{m'' m'}(\vbh)
}
\esub
%---------------------------
to return to the lab frame. Using the identity from Eq.~\eqref{eq:D_using_Ylm}
%%---------------------------
%\bealf{
%D^{\ell}_{mm'}(\vbh)&=(-1)^{m'}\,\sqrt{\frac{4\pi}{2\ell+1}}\,{}_{-m'}Y^*_{\ell m}(\vbh)\,\expf{-i m' \chi}
%}
%%---------------------------
we find
%---------------------------
\bsub
\label{eq:final_general_expressions_S_final}
\bealf{
\left.\frac{\id  n^i(\nu,\vgh)}{ \id \tau}\right|_{\rm T}
&=\sum_{\ell m m'' \ell'' j}
{_{s^i}}Y_{\ell m}(\vgh)\,
{_{ij}}\hat{\mathcal{S}}^{m m''}_{\ell \ell''}(\nu, \vb)
\,n_{\ell'' m''}^j(\nu)
\\
{_{ij}}\hat{\mathcal{S}}^{m m''}_{\ell \ell''}(\nu, \vb)&=
\sum_{m'}
\frac{4\pi\,{}_{-m'}Y^*_{\ell m}(\vbh)\,{_{ij}}\hat{\mathcal{S}}^{m'}_{\ell \ell''}(\nu, \beta)\,{}_{-m'}Y_{\ell'' m''}(\vbh)}{\sqrt{(2\ell+1)(2\ell''+1)}}
}
\esub
%---------------------------
for general direction of the boost. This expression includes all orders in $\varv/c$ and will be extremely useful in our derivations below. We note that this is also consistent with the notion of a scattering operator transformation $\mathcal{\hat{S}}=\hat{R} \mathcal{\hat{S}}' \hat{R}^\dagger$, where $\mathcal{\hat{S}}'$ is the scattering operator in the frame special frame $S'$ and we rotate by $\hat{R}^{-1}=\hat{R}^\dagger$.

\vspace{-3mm}
\subsection{Scattering by isotropic electron distributions}
\label{sec:scat_iso_matrix}
%----------------------------
It is clear that anisotropic electron distributions will generate polarization at second order in $\beta$ even in the absence of polarization before the scattering (see Sec.~\ref{sec:rel_pSZ}). However, in this section we shall consider what happens when the intrinsic electron distribution is isotropic. Without any photon anisotropies, we naturally expect no polarization to be generated and also no anisotropies to be generated in this way, but we can demonstrate even more general aspects using the results from the boost operator approach. 

Assuming that the electron momentum distribution is isotropic, we can average ${_{ij}}\hat{\mathcal{S}}^{m m''}_{\ell \ell''}(\nu, \vb)$ in Eq.~\eqref{eq:final_general_expressions_S_final} over $\vbh$. This then yields
%---------------------------
\bealf{
\label{eq:S_final_iso}
&\left<{_{ij}}\hat{\mathcal{S}}^{m m''}_{\ell \ell''}(\nu, \vb)\right>_{\vbh}
=\int \frac{\id \vbh}{4\pi}\,{_{ij}}\hat{\mathcal{S}}^{m m''}_{\ell \ell''}(\nu, \vb)
=
\frac{\delta_{\ell \ell''}\,
\delta_{m m''}}{2\ell+1}
\sum_{m'}{_{ij}}\hat{\mathcal{S}}^{m'}_{\ell \ell}(\nu, \beta)
\nonumber\\
&\;\,=
\delta_{\ell \ell''}\,
\delta_{m m''}
\left\{
\frac{\Dbo{0 0}{\ell 0 \ell}{0}(\beta)}{2\ell+1}\,
\delta^{i0}\delta^{j0}
-\delta^{ij}+\sum_{m'=-2}^{2}
\frac{c^{ij}}{10}\,\frac{\Dbo{i j}{\ell 2 \ell}{m'}(\beta)}{2\ell+1}
\right\}
}
%---------------------------
for the averaged scattering operator, allowing us to directly read off the evolution equation for each $n_{\ell m}^i$ from Eq.~\eqref{eq:final_general_expressions_S_final} once averaged over the electron momentum distribution, $f(p)$. This form of the operator shows that for isotropic electron distributions Thomson scattering does not couple photons between different multipoles $(\ell,m)$ and $(\ell',m')$. The scattering operator is furthermore independent of $m$ but in principle allows the mixing of polarization and intensity terms for $\ell\geq 2$. This also means that if no anisotropy at a given $(\ell, m)$ is present we will not be able to generate any new anisotropies.

For scattering of the CMB monopole, we obtain the usual relativistic thermal SZ effect \citep[cf.][]{ChlubaBO25}. We now also include higher temperature anisotropies and polarization states for further illustration. We start the discussion by calculating the properties of the Doppler operator in more detail.
Using the symmetries of the boost operator \citep{ChlubaBO25}, we can show that 
%---------------------------
\bealf{
\label{eq:Sym_D}
\Dbo{i j}{\ell \ell' \ell''}{0}
&=\frac{
\Bb{-1}{-s^i}{0}{\ell\ell'}(-\beta)
\,
\Bb{0}{-s^j}{0}{\ell'\ell''}(\beta)}{\gamma}
\equiv
\frac{
\Bb{-1}{s^i}{0}{\ell\ell'}(-\beta)
\,
\Bb{0}{s^j}{0}{\ell'\ell''}(\beta)}{\gamma}
}
%---------------------------
is independent of the signs of the spin-weights involved. With $\Bb{d}{s^j}{m}{\ell\ell'}(\beta)=\Bb{d}{-s^j}{-m}{\ell\ell'}(\beta)$, we then have
%---------------------------
\bealf{
\label{eq:Sym_D}
&\Dbo{i j}{\ell \ell' \ell''}{}\equiv\sum_{m'} \Dbo{i j}{\ell \ell' \ell''}{m'}
\\[-1mm]
\nonumber 
&\quad\;\,
=\Dbo{i j}{\ell \ell' \ell''}{0}
 +
\sum_{m'>0} 
\frac{\Bb{-1}{-s^i}{m'}{\ell\ell'}(-\beta)\,\Bb{0}{-s^j}{m'}{\ell'\ell''}(\beta)+\Bb{-1}{s^i}{m'}{\ell\ell'}(-\beta)\,\Bb{0}{s^j}{m'}{\ell'\ell''}( \beta)}{\gamma}\,
}
%---------------------------
meaning that also $\Dbo{i j}{\ell \ell' \ell''}{}$ is independent of the signs of the involved spin-weights. This reduces the computational burden significantly. Note that the sum over $m'$ is determined by the lowest multipole that is involved.

We next compute all required $m$-averaged operators. For this we follow the procedure as explained in Appendix~\ref{app:Kernel_computation}, again easing the computation. Up to fourth order in $p$ all operators are explicitly shown in Appendix~\ref{app:Doppler_op_p4}. However, we can simplify the expressions by realizing that the {\it diffusion operator}, $\oDnu=\nu^{-2}\partial_\nu \nu^4 \partial_\nu$ can be written as $\oDnu=\oOnu^2-3\oOnu$ \citep{chluba_spectro-spatial_2023-I}. Similarly, we can write $\oDnu^2=\oOnu^4-6\oOnu^3+9\oOnu^2$. Using these replacements, and defining ${_{ij}}\hat{\mathcal{S}}_{\ell}(\nu, \beta)=\left<{_{ij}}\hat{\mathcal{S}}^{m m''}_{\ell \ell''}(\nu, \vb)\right>_{\vbh}$, for all intensity states we find 
\begin{comment}
%---------------------------
\bsub
\label{eq:D_explicit}
\bealf{
\Dbo{0 0}{000}{}&\approx 1+\oDnu\frac{p^2}{3} 
-\frac{8}{3}\left(\oDnu-\frac{\oDnu^2}{4}\right)\frac{p^4}{15} 
\nonumber\\
\frac{\Dbo{0 0}{101}{}}{3}&\approx -\frac{2}{3}\left(1+\frac{\oDnu}{2}\right)\frac{p^2}{3} 
+\frac{8}{3}\left(1+\frac{1}{4}\left[\oDnu-\frac{\oDnu^2}{2}\right]\right)\frac{p^4}{15} 
\nonumber\\
\frac{\Dbo{0 0}{202}{}}{5}&\approx \frac{2}{15}\left(\oDnu+\frac{\oDnu^2}{2}\right)\frac{p^4}{15} 
\\[2mm]
\frac{\Dbo{0 0}{020}{}}{10}&\approx -\frac{2}{15}\left(\oDnu-\frac{\oDnu^2}{4}\right)\frac{p^4}{15}
\nonumber\\
\frac{\Dbo{0 0}{121}{}}{30}&\approx \frac{4}{15}\left(1-\frac{\oDnu}{4}\right)\frac{p^2}{3}-\left(\frac{28}{15}-\frac{11\oDnu}{15}+\frac{\oDnu^2}{15}\right)\frac{p^4}{15}
\nonumber\\
\frac{\Dbo{0 0}{222}{}}{50}&\approx \frac{1}{10}-\frac{3}{5}\left(1-\frac{\oDnu}{6}\right)\frac{p^2}{3}
+\left(\frac{144}{35}-\frac{17\oDnu}{15}+\frac{8\oDnu^2}{105}\right)\frac{p^4}{15} 
\nonumber\\
\frac{\Dbo{0 0}{323}{}}{70}&\approx \frac{6}{35}\left(1-\frac{\oDnu}{4}\right)\frac{p^2}{3}
-\left(\frac{72}{35}-\frac{24\oDnu}{35}+\frac{3\oDnu^2}{70}\right)\frac{p^4}{15} 
\nonumber\\
\frac{\Dbo{0 0}{424}{}}{90}&\approx \left(\frac{8}{21}-\frac{2}{15}\left[\oDnu-\frac{\oDnu^2}{14}\right]\right)\frac{p^4}{15}. 
}
\esub
%---------------------------
\end{comment}
%---------------------------
\bealf{
\label{eq:S_explicit}
{_{00}}\hat{\mathcal{S}}_{0}&\approx \oDnu\frac{p^2}{3} 
-\frac{7\oDnu}{10}(4-\oDnu)\frac{p^4}{15} 
\nonumber\\
1+{_{00}}\hat{\mathcal{S}}_{1}&\approx-\frac{2}{5}(1+ \oDnu)\frac{p^2}{3} 
+\frac{2}{5}\left(2+\frac{7\oDnu}{2}-\oDnu^2\right)\frac{p^4}{15} 
\nonumber\\
1+{_{00}}\hat{\mathcal{S}}_{2}&\approx\frac{1}{10} -\frac{1}{10}(6-\oDnu)\frac{p^2}{3} +
\left(\frac{144}{35}-\oDnu+\frac{\oDnu^2}{7}\right)\frac{p^4}{15}
\nonumber\\
1+{_{00}}\hat{\mathcal{S}}_{3}&\approx \frac{3}{70}(4-\oDnu)\frac{p^2}{3}
-\frac{3}{70}(12-\oDnu)(4-\oDnu)\frac{p^4}{15} 
\nonumber\\
1+{_{00}}\hat{\mathcal{S}}_{4}&\approx \frac{1}{105}(12-\oDnu)(4-\oDnu)\frac{p^4}{15}. 
}
%---------------------------
We note the leading order diagonal contribution as $-1$, accounting for the scattering of photons out of the line of sight.

For the operators involving one polarization state we find
%---------------------------
\bealf{
\label{eq:S_explicit_s0}
1+{_{\pm 0}}\hat{\mathcal{S}}_{2}&\approx \frac{1}{10}-\frac{1}{10}(4-\oDnu)\frac{p^2}{3}
+\frac{2}{35}(10-\oDnu)(4-\oDnu)\frac{p^4}{15} 
\nonumber\\
1+{_{\pm 0}}\hat{\mathcal{S}}_{3}&\approx \frac{1}{14\sqrt{5}}(4-\oDnu)\frac{p^2}{3}
-\frac{1}{14\sqrt{5}}(10-\oDnu)(4-\oDnu)\frac{p^4}{15}
\nonumber\\
1+{_{\pm 0}}\hat{\mathcal{S}}_{4}&\approx \frac{1}{42\sqrt{15}}(10-\oDnu)(4-\oDnu)\frac{p^4}{15} 
\nonumber\\
{_{0\pm}}\hat{\mathcal{S}}_{\ell}
&\equiv {_{\pm 0}}\hat{\mathcal{S}}_{\ell}.
}
%---------------------------
We were unable to explicitly prove the last identity but we confirmed it to be correct even for higher order terms. 

Finally, for the operators connecting polarization states we have 
%---------------------------
\bsub
\label{eq:S_explicit_psps}
\bealf{
1+{_{\pm\pm}}\hat{\mathcal{S}}_{2}
&\approx 
\frac{1}{10}-\frac{1}{30}(10-\oDnu)\frac{p^2}{3}
+\frac{1}{105}(18-\oDnu)(10-\oDnu)\frac{p^4}{15}
\nonumber\\
1+{_{\pm\pm}}\hat{\mathcal{S}}_{3}
&\approx \frac{1}{42}(4-\oDnu)\frac{p^2}{3}
-\frac{1}{84}(18-\oDnu)(4-\oDnu)\frac{p^4}{15} 
\nonumber\\
1+{_{\pm\pm}}\hat{\mathcal{S}}_{4}
&\approx \frac{1}{252}(10-\oDnu)(4-\oDnu)\frac{p^4}{15} 
\\[2mm]
{_{\pm\mp}}\hat{\mathcal{S}}_{2}
&\approx {_{\pm\pm}}\hat{\mathcal{S}}_{2}+\frac{2}{15}(2+\oDnu)\frac{p^2}{3} 
-\frac{2}{15}(5-\oDnu)(2+\oDnu)\frac{p^4}{15} 
\nonumber\\
{_{\pm\mp}}\hat{\mathcal{S}}_{3}&\approx {_{\pm\pm}}\hat{\mathcal{S}}_{4}
+\frac{1}{42}(4-\oDnu)(2+\oDnu)\frac{p^4}{15}
\nonumber\\
{_{\pm\mp}}\hat{\mathcal{S}}_{4}&\approx {_{\pm\pm}}\hat{\mathcal{S}}_{4}
}
\esub
%---------------------------
What do the various terms mean physically? First and foremost, powers of the diffusion operator $\oDnu$ do not change the number of photons but just redistribute them in energy, leading to spectral evolution of the photon distribution. 
Next, all operators connecting the same initial and final spin-weight state simply lead to an evolution of the related spectrum and photon occupation of that multipole. The time-scale for this evolution furthermore depends on $\ell$. %Those that mix spin-weight states cause changes of the polarization degree and the polarization direction. 

However, a second aspect comes to light: all the required operators are linear combinations of powers of the diffusion operator. In addition, significant structure is visible in the related coefficients, hinting at another way to simplify matters, a problem the we will leave to future work. We now collect all terms up to first order in the electron temperature for further applications.

\subsubsection{Leading order terms ($p=0$)}
%---------------------------
Let us first set $p=0$. This evidently reduces the equation to the standard Thomson scattering terms which lead to the isotropization of the medium without changing the total number of photons: 
%---------------------------
\bealf{
\label{eq:ndot_p0}
\left.\frac{\id  n^i_{\ell m}(\nu)}{\id \tau}\right|^{p=0}_{\rm T}
&=\sum_{j}
\left\{\delta_{i0}\delta_{j0}\delta_{\ell 0}+\frac{c_{ij}\delta_{\ell 2}}{10}-\delta_{ij}\right\} 
\,n_{\ell m}^j(\nu).
}
%---------------------------
Specifically, there is no change of the photon monopole $n^0_{00}$ by Thomson scattering while all isotropies damp as $-\dot\tau$ from the diagonal element of the averaged scattering operator. For the quadrupole, in addition one has contributions from $\sum_j \frac{c_{ij}}{10} n_{2 m}^j(\nu)$ that lead to mixing between the spin-weight states. Note that due to the form of $c_{ij}$, $n^0_{2m}$ sources equal amounts of $n^\pm_{2m}$ and similarly for the reverse. 

\subsubsection{Aberration-induced temperature corrections}
%---------------------------
Next we consider only those terms without a diffusion operator. Those describe the temperature corrections of the Thomson scattering rate (no change of the photon energy) from light aberration effects. Replacing $p^2$ by the thermal average $3\The=3 k\Te/\me c^2$ and dropping higher order terms, for the intensity states ($i=j=0$) this means the new terms 
%---------------------------
$$\frac{\id  n^{0}_{\ell m}(\nu)}{\id \tau}\approx \The\left(-\frac{2}{5} \delta_{\ell 1}-\frac{3}{5} \delta_{\ell 2}+\frac{6}{35} \delta_{\ell 3}\right)\,n_{\ell m}^0(\nu)$$
%---------------------------
as also obtained in \citet{Chluba2012}. These terms lead to a temperature-dependent correction to the intensity scattering rates of the dipole through octupole \citep{chluba_spectro-spatial_2023-II}.
For $\ell=2, 3$, additional terms appear when considering the polarization states. 
To capture those temperature dependent corrections, we therefore introduce the matrices $a^{(2)}_{ij}$ and $a^{(3)}_{ij}$\footnote{For $\bm{a}^{(2)}$ we scaled by a factor of $-\frac{3}{5}\The$ and for $\bm{a}^{(3)}$, we pulled out the factor of $\frac{6}{35}\The$ to normalize everything to the intensity scattering rate.} 
%---------------------------
\bsub
\bealf{
\bm{a}^{(2)} &= 
 \begin{pmatrix}
    1 & -\sqrt{\frac{8}{3}}\frac{1}{2} &-\sqrt{\frac{8}{3}}\frac{1}{2}
    \\
    -\sqrt{\frac{8}{3}} & \frac{5}{3} & \frac{1}{3}
    \\
    -\sqrt{\frac{8}{3}} & \frac{1}{3} & \frac{5}{3}
    \end{pmatrix}
\\
\bm{a}^{(3)} &= 
    \begin{pmatrix}
    1 &-\sqrt{\frac{10}{3}}\frac{1}{2} &-\sqrt{\frac{10}{3}}\frac{1}{2}
    \\
    -\sqrt{\frac{10}{3}}  & \frac{5}{3} & \frac{5}{3}
    \\
    -\sqrt{\frac{10}{3}}  & \frac{5}{3} & \frac{5}{3} 
    \end{pmatrix},
}
\esub
%---------------------------
which yield the new terms
%---------------------------
\bealf{
\label{eq:ndot_p0_The}
\left.\frac{\id  n^i_{\ell m}}{\id \tau}\right|^{\The}_{\rm T}
&=\sum_{j}
\The\left\{-\frac{2}{5}\delta_{i0}\delta_{j0}\delta_{\ell 1}-\frac{3a^{(2)}_{ij}\delta_{\ell 2}}{5}+\frac{6 a^{(3)}_{ij}\delta_{\ell 3}}{35}\right\} 
 n_{\ell m}^j.
}
%---------------------------
The forms of $\bm{a}^{(2)}$ and $\bm{a}^{(3)}$ reveal some interesting physical aspects: while $\bm{a}^{(3)}$ has a similar structure to $\bm{c}$, thereby leading to a sourcing and state-mixing correction for multipoles with $\ell=3$, we find that with temperature corrections $\bm{a}^{(2)}$ reveal an additional asymmetry in the mixing between the $n^+$ and $n^-$ states, modifying how polarization states mix via scattering. The sourcing by intensity and also damping into intensity states is again the same for $n^+$ and $n^-$. 

For the effects to be noticeable, one requires a sufficient Compton $y$-parameter $y_{\rm e}=\int \The \id \tau$, as for SZ clusters, possibly leading to a $\simeq 1\%$ correction relative to the transport without the aberration-induced temperature corrections \citep{Sazonov1999, Challinor2000, Itoh2000Pol}.

\subsubsection{Including leading order Doppler diffusion corrections}
%---------------------------
Finally, we write the diffusion terms at leading order in the temperature. Defining the  diffusion matrices
%---------------------------
\bealf{
\bm{d}^{(2)} &= 
    \begin{pmatrix}
    1 &-\frac{\sqrt{6}}{2} &-\frac{\sqrt{6}}{2}
    \\
    -\sqrt{6} & 1 & 5
    \\
    -\sqrt{6} & 5 & 1
    \end{pmatrix}  
    \qquad \text{and} \qquad
\bm{d}^{(3)}=\bm{a}^{(3)}    
}
%---------------------------
this allows us to compactly write all the diffusion terms as
%---------------------------
\bealf{
\label{eq:ndot_p0_The_Diff}
\left.\frac{\id  n^i_{\ell m}}{\id \tau}\right|^{\rm D}_{\rm T}
&=\sum_{j}
\The \left\{\delta_{i0}\delta_{j0}\left(\delta_{\ell 0}-\frac{2}{5}\delta_{\ell 1}\right)+\frac{d^{(2)}_{ij}\delta_{\ell 2}}{10}-\frac{3 d^{(3)}_{ij} \delta_{\ell 3}}{70}
\right\} 
\oDnu n_{\ell m}^j.
}
%---------------------------
This expression shows that for $\ell=2$ also the diffusion corrections lead to a difference in the spin-state mixing terms, causing a frequency-dependent correction to the polarization fraction. The terms for $i=j=0$ are consistent with those given in, e.g., \citet{Hu1994pert, Bartolo2006, Chluba2012}

\subsection{Moving cloud of thermal electrons}
%----------------------------
We next consider the scattering operator for a moving cloud of relativistic thermal electrons. This case will be of interest for applications to the relativistic polarized SZ effect and accounts for kinematic corrections to the scattering process (Sec.~\ref{sec:rel_pSZ}). 

The relevant electron distribution function can be expressed as \citep{Lee2024SZpack, ChlubaBO25b} 
%------------------------------
\bsub
\label{eq:moving_ele}
\bealf{
f_{\rm p}(\vek{p}) 
&=
\frac{(N_{\rm e}/\gammac)\, \exp\left(-\frac{\gammac\gamma}{\The}\right)}{4\pi \The K_2(1/\The) }
\times \exp\left(\frac{\vek{p}_{\rm p} \cdot \vek{p}}{\The}\right)
\\
\exp\left(\frac{\vek{p}_{\rm p} \cdot \vek{p}}{\The}\right)&=\sqrt{\frac{\pi\,\The}{2 p_{\rm p} p}}\sum_{L=0}^\infty (2L+1) \,I_{L+\frac{1}{2}}\left(\frac{p_{\rm p} p}{\The}\right) P_L(\hat{\vek{p}}_{\rm p} \cdot \vbh),
}
\esub
%------------------------------
where the cluster's peculiar motion is in the direction $\vbh_{\rm p}$, with a speed $\beta_{\rm p}=p_{\rm p}/\gamma_{\rm p}$ and momentum $\vek{p}_{\rm p}$. Also, $I_n(x)$ and $K_n(x)$ are the modified Bessel function of the first and second kind, respectively. We furthermore use the dimensionless temperature $\The = k\Te/\me c^2$ and normalized the distribution such that $\int f_{\rm p}(\vek{p})\id^3 p =N_{\rm e}$, where $N_{\rm e}$ is the lab frame electron number density. 

To simplify matters, we can use the addition theorem to rewrite $P_{L}(\hat{\vek{p}}_{\rm p} \cdot \vbh)$ and thus introduce
%------------------------------
\bealf{
\label{eq:moving_ele_rewrite}
f_{\rm p}(\vek{p}) 
&=\sum_{LM}\,f_{L}(\gamma, \gamma_{\rm p})\,Y^*_{L M}(\hat{\vek{p}}_{\rm p})Y_{L M}(\vbh)
\,
\\
f_{L}(\gamma, \gamma_{\rm p}) 
&=
\frac{\exp\left(-\frac{\gammac\gamma}{\The}\right)}{\gammac \The K_2(1/\The) }\,\sqrt{\frac{\pi\,\The}{2 p_{\rm p} p}}\,I_{L+\frac{1}{2}}\left(\frac{p_{\rm p} p}{\The}\right)
}
%------------------------------
We next consider the average of Eq.~\eqref{eq:final_general_expressions_S_final} over the directions of $\id \vb$; unlike in Eq.~\eqref{eq:S_final_iso} we have dropped the assumption of an isotropic distribution of electron momenta. Focusing on the angular dependence only, we then encounter the integrals
%------------------------------
\bealf{
\label{eq:betahat_int_gen}
{}_{m'}\mathcal{I}^{\ell,\ell'',L}_{-m,m'',M}
=
&\int \id \vbh \, \frac{4\pi\, {}_{-m'}Y^*_{\ell m}(\vbh)\,{}_{-m'}Y_{\ell'' m''}(\vbh)\,Y_{L M}(\vbh)}{\sqrt{(2\ell+1)(2\ell''+1)}}
\nonumber \\
&=\frac{4\pi \,(-1)^{m'+m} {}_{m',-m',0}\mathcal{G}^{\ell,\ell'',L}_{-m,m'',M}}{\sqrt{(2\ell+1)(2\ell''+1)}},
}
%------------------------------
where we introduced the Gaunt integral (see Appendix~\ref{app:Gaunt}). We then find the general solution
%---------------------------
\bsub
\label{eq:final_general_expressions_S_final_kin}
\bealf{
&\left.\frac{\id  n^i(\nu,\vgh, \vek{p}_{\rm p})}{ \id \tau}\right|_{\rm T}
=\sum_{\ell m m'' \ell'' j}
\!\!{_{s^i}}Y_{\ell m}(\vgh)\,
{{}^{\rm th}_{ij}}\hat{\mathcal{S}}^{m m''}_{\ell \ell''}(\nu, \vek{p}_{\rm p})
\,n_{\ell'' m''}^j(\nu)
\\
&{{}^{\rm th}_{ij}}\hat{\mathcal{S}}^{m m''}_{\ell \ell''}(\nu,\vek{p}_{\rm p})=\sum_{LM} 
Y^*_{L M}(\hat{\vek{p}}_{\rm p})\,
{}^{\rm th}\hat{\mathcal{S}}^{m m''M}_{\ell \ell''L}(\nu, p_{\rm p})
\\
&{{}^{\rm th}_{ij}}\hat{\mathcal{S}}^{m m''M}_{\ell \ell''L}(\nu, p_{\rm p})=\int 
p^2 \id p \, f_{L}(\gamma, \gamma_{\rm p})\,
{_{ij}}\hat{\mathcal{S}}^{m m''M}_{\ell \ell''L}(\nu, \beta=p/\gamma)
\\
&{_{ij}}\hat{\mathcal{S}}^{m m'' M}_{\ell \ell'' L}(\nu, \beta)=
\sum_{m'}
{_{ij}}\hat{\mathcal{S}}^{m'}_{\ell \ell''}(\nu, \beta)\,
{}_{m'}\mathcal{I}^{\ell,\ell'',L}_{-m,m'',M}.
}
\esub
%---------------------------
These expressions become more tractable when considering specific examples as below. They capture all temperature and kinematic corrections to the scattering operator in the Thomson limit. Using the Doppler operators, one can generate term by term in $\beta_{\rm p}$ and $\The$.

\section{Applications to CMB}
\label{sec:cmb_apps}
%----------------------------
In this section we now consider some illustrative examples in connection to the scattering of CMB photons. We first start with the scattering of the CMB monopole blackbody and then consider the scattering of CMB anisotropies, which allows us to illustrate the generation of polarization by scattering.

\subsection{Scattering of the average CMB monopole}
\label{sec:rel_pSZ}
%----------------------------
To illustrate how the generation of polarization due to scattering off moving electrons works, let us start with the simplest case where we only consider the scattering of the CMB intensity monopole, with occupation number $n^0=\nbb=1/(\expf{x}-1)$. Since in this case $n_{00}=\sqrt{4\pi} \nbb$, we can set $\ell''=m''=0$ in the expression Eq.~\eqref{eq:final_general_expressions_S_final_kin}. This means ${}_{m'}\mathcal{I}^{\ell,0,L}_{-m,0,M}=\sqrt{\frac{4\pi}{2\ell+1}}{}\delta_{m'0}\delta_{mM}\delta_{\ell L}$. The required scattering operators are then
%---------------------------
\bealf{
\label{eq:monopole_scattering}
{_{i0}}\hat{\mathcal{S}}^{0}_{\ell 0}(\nu, \beta)
&=\Dbo{0 0}{\ell 0 0}{0}\,
\delta^{i0}
+\frac{c^{i0}}{10}\,\Dbo{i 0}{\ell 2 0}{0}-\delta_{\ell0}\delta^{i0}
+\frac{\beta}{\sqrt{3}} \,\delta_{\ell 1}
\delta^{i0}.
}
%---------------------------
The terms for $i=0$ reproduce the result from \citet{ChlubaBO25b} relating to the relativistic SZ effect with all kinematic corrections \citep[see Eq.~(2) of][]{ChlubaBO25b}.
For $i=\pm$, we find the compact expression
%---------------------------
\bsub
\label{eq:monopole_scattering_pm}
\bealf{
&\left.\frac{\id  n^\pm(\nu,\vgh)}{ \id \tau}\right|_{\rm T}
=
\sum_{\ell m}
\frac{4\pi \,{_{\pm 2}}Y_{\ell m}(\vgh) \,Y^*_{\ell m}(\hat{\vek{p}}_{\rm p})}{2\ell+1}
\,\mathcal{\hat{S}}^\pm_{\ell}(\nu, \The,\beta_{\rm p})\,\nbb(\nu)
\\[-1mm]
&\mathcal{\hat{S}}^\pm_{\ell}(\nu, \The, \beta_{\rm p})=\int_0^\infty
p^2
f_\ell(\gamma, \gamma_{\rm p})
\,\mathcal{\hat{S}}^\pm_{\ell}(\nu, p)\,
\text{d}p
\\ \nonumber 
&\mathcal{\hat{S}}^\pm_{\ell}(\nu, \beta)\equiv \sqrt{2\ell+1}\,{_{\pm, 0}}\hat{\mathcal{S}}^{0}_{\ell 0}(\nu, \beta)
=-\frac{\sqrt{6}}{10}\,\sqrt{2\ell+1}\,\Dbo{\pm, 0}{\ell 2 0}{0}(\nu, \beta).
}
\esub
%---------------------------
in the general frame. Looking at the definition of $\Dbo{ij}{\ell \ell' \ell''}{0}(\nu, \beta)$, and using the results of \citet{ChlubaBO25b}, the only new boost operator elements that we have to work out are $\Bbob{\pm2}{\ell2}{0}(\nu, -\beta)$, which can be computed as explained in the Appendix~\ref{app:Kernel_raise}.

To write the final relativistic polarized SZ expressions, (using Appendix~\ref{app:addition_sYlm}) we introduce
%------------------------------
\bealf{
\label{eq:identitiy_spin}
\mathcal{P}^\pm_\ell(\vgh,\hat{\vek{p}}_{\rm p})
&=\sum_{m} \frac{4\pi\,{_{\pm 2}} Y_{\ell m}(\vgh)\, Y^*_{\ell m}(\hat{\vek{p}}_{\rm p})}{2\ell+1}
\nonumber\\
&\equiv \sqrt{\frac{4\pi}{2\ell+1}}\,{_{\pm 2}} Y_{\ell 0}(\theta, \alpha)\,\expf{\mp 2 i \chi},
}
%------------------------------
where $\cos\theta=\vgh\cdot\vbh_{\rm p}$, $\alpha$ is the angle that $\vgh-\vbh_{\rm p}$ makes with $\hat{\vek{e}}_{\theta}(\vbh_{\rm p})$, and $\chi$ is the angle that $\vgh-\vbh_{\rm p}$ makes with $\hat{\vek{e}}_{\theta}(\vgh)$ \citep[cf. Eq.~(7) and Fig.~1 of][]{Hu1997}. The angle $\alpha$ does not affect the result and $\chi$ can be thought of as the angle that the transverse velocity makes with respect to sky projection. For $\chi=0$, the transverse velocity is aligned with $\hat{\vek{e}}_{\theta}(\vgh)$ and the polarization direction is along $\hat{\vek{e}}_{\phi}(\vgh)$ implying a pure $Q$ polarization with no $U$ contribution.

Collecting terms, we finally obtain
%---------------------------
\bsub
\label{eq:monopole_scattering_pm_final}
\bealf{
\Delta n^\pm(\nu, \The, \vgh, \vb_{\rm p})\big|_{\rm pSZ}
&=
\tau^*\,\hat{\mathcal{S}}^\pm_{\rm pSZ}(\nu, \The, \vgh, \vb_{\rm p})\,\nbb(\nu)
\\[0.5mm]
\hat{\mathcal{S}}^\pm_{\rm pSZ}(\nu, \The, \vgh, \vb_{\rm p})&=
\sum_{\ell}
\mathcal{\hat{S}}^\pm_{\ell}(\nu, \The, \beta_{\rm p})\,\mathcal{P}^\pm_\ell(\vgh, \hat{\vb}_{\rm p})
}
\esub
%---------------------------
where $\tau^*\equiv \tau/[1-\vb_{\rm p}\cdot \vgh]$ with lab-frame optical depth $\tau=\int \Ne \sigma_{\rm T} \id l$ along the line-of-sight \citep[e.g.,][]{Chluba2012SZpack}. 

Equation~\eqref{eq:monopole_scattering_pm_final} is the final result for the relativistic polarized SZ effect at all orders in $\beta_{\rm p}$ and the electron temperature. The remaining integral over the electron momenta can be computed term by term in orders of the $\beta_{\rm p}$ and $\The$ using the same methods as described in \citet{ChlubaBO25b}. This task can be easily carried out using {\tt Mathematica}. We now give the results for limiting cases and compare to some of the existing expressions in the literature \citep[e.g.,][]{Itoh2000Pol, Challinor2000}. To show the potential of the operator method we also give some of the higher order terms for illustration, noting that higher order terms can be easily generated.

\subsubsection{Zero temperature limit}
%---------------------------
To obtain the pure kinematic relativistic SZ effect without any temperature corrections but at all orders in $\beta_{\rm p}$, we can directly use Eq.~\eqref{eq:final_general_expressions_S_final} by inserting $\vb=\vb_{\rm p}$ and recognizing the connection with $\mathcal{P}^\pm_2(\vgh, \hat{\vb}_{\rm p})$ in Eq.~\eqref{eq:identitiy_spin}. The same result can be obtained by taking the zero temperature limit of Eq.~\eqref{eq:monopole_scattering_pm_final}. 

Let us first consider the terms relating to $\ell=2$. For these we have (see Appendix~\ref{app:spin_harmonics} and following \ref{app:Kernel_raise})
%---------------------------
\bealf{
\mathcal{P}^\pm_2(\vgh, \hat{\vb}_{\rm p})&=\sqrt{\frac{3}{8}}\,(1-\mu_{\rm p}^2)\,\expf{\mp 2 i \chi}
\nonumber \\
\Dbo{\pm, 0}{2 2 0}{0}(\nu, \beta_{\rm p})
&\approx \frac{\oOnu(1+\oOnu)}{3\sqrt{5}}\,\beta^2_{\rm p}\Bigg[1
-
\left(\frac{3}{7}+\frac{3\oOnu}{7}-\frac{\oOnu^2}{7}\right)\beta^2_{\rm p}
\\ \nonumber
&\qquad-\left(\frac{11}{49}+\frac{2\oOnu}{21}-\frac{50\oOnu^2}{441}+\frac{8\oOnu^3}{147}-\frac{4\oOnu^3}{441}\right)\beta^4_{\rm p}
\Bigg]
}
%---------------------------
up to sixth order in $\beta_{\rm p}$ and where $\oOnu=-\nu \partial_\nu$. Defining $x=h\nu/\kB\Tg$ and since $\oOnu(1+\oOnu)=\oOx{2}$, we find the leading order term
%---------------------------
\bealf{
\label{eq:monopole_scattering_pSZ}
\Delta n^\pm\big|_{\rm pSZ}
&\approx
-\frac{\tau^* \beta^2_{\rm p, \perp}}{20}\,\frac{x^2 \expf{x} (\expf{x}+1)}{(\expf{x}-1)^3}\,\expf{\mp 2 i \chi},
}
%---------------------------
which is the classical result for the polarized SZ effect \citep{Sunyaev1980, Sazonov1999, Itoh2000Pol} with the transverse velocity component, with $\beta_{\rm p, \perp}=\beta_{\rm p} (1-\mu_{\rm p}^2)^{1/2}$. We note that in terms of $n^Q$ and $n^U$, we have %---------------------------
\bsub
\label{eq:monopole_scattering_pSZ_QU}
\bealf{
n^Q&=-\frac{\tau^* \beta^2_{\rm p, \perp}}{20}\,\frac{x^2 \expf{x} (\expf{x}+1)}{(\expf{x}-1)^3}\,\cos (2\chi)
\\
n^U&=-\frac{\tau^* \beta^2_{\rm p, \perp}}{20}\,\frac{x^2 \expf{x} (\expf{x}+1)}{(\expf{x}-1)^3}\,\sin (2\chi)
}
\esub
%---------------------------
in the considered case. The scattered light is thus polarized perpendicular to the direction of $\vb_{\rm p,\perp}$. For $\chi=0$, we  have $n^Q<0$ and $n^U=0$ with $\vb_{\rm p,\perp}\parallel \hat{\vek{e}}_{\theta}(\vgh)$. Since $n^Q=n_\parallel-n_\perp$ with $n_\parallel$ being the photon occupation number in the direction $\hat{\vek{e}}_{\theta}(\vgh)$ and $n_\perp$ corresponding in the direction $\hat{\vek{e}}_{\phi}(\vgh)$, the overall minus sign signifies that $n_\parallel=0$ and $n^Q$ is in the direction of $\hat{\vek{e}}_{\phi}(\vgh)$ \citep[cf. also][]{Itoh2000Pol}.

Including terms up to sixth order in $\beta_{\rm p}$, we can write
%---------------------------
\bealf{
\label{eq:monopole_scattering_pSZ_sixth}
&\Delta n^\pm\big|^{\ell=2}_{\rm pSZ}
\approx
-\frac{\tau^* \beta^2_{\rm p, \perp}\,\expf{\mp 2 i \chi}}{20}\,
\Bigg[
\oOx{2}+\left(\oOx{2}+\frac{8\oOx{3}}{7}+\frac{\oOx{4}}{7}\right)\beta^2_{\rm p}
\\ \nonumber
&\qquad+\left(\oOx{2}+\frac{16\oOx{3}}{7}+\frac{54\oOx{4}}{49}+\frac{80\oOx{5}}{441}+\frac{4\oOx{6}}{441}\right)\beta^4_{\rm p}
\Bigg]\,\nbb(x)
}
%---------------------------
in the zero temperature limit. Note there is no term $\mathcal{O}(\beta^3_{\rm p, \perp})$, which only appears when including terms for $\ell=3$:
%---------------------------
\bealf{
\mathcal{P}^\pm_3(\vgh, \hat{\vb}_{\rm p})&=\sqrt{\frac{15}{8}}\,\mu_{\rm p}(1-\mu_{\rm p}^2)\,\expf{\mp 2 i \chi}
\\ \nonumber
\Dbo{\pm, 0}{3 2 0}{0}(\nu, \beta_{\rm p})
&\approx \frac{\oOnu(1+\oOnu)^2}{3\sqrt{35}}\,\beta^3_{\rm p}\Bigg[1
-
\left(\frac{5}{21}+\frac{23\oOnu}{63}-\frac{8\oOnu^2}{63}\right)\beta^2_{\rm p}
\Bigg].
}
%---------------------------
Here we computed the required boost operator $\Bbob{\pm2}{32}{0}(\nu, -\beta)$ using Eq.~\eqref{eq:raise_spin2}. This then gives
%---------------------------
\bealf{
\label{eq:monopole_scattering_pSZ_sixth_3}
\Delta n^\pm\big|^{\ell=3}_{\rm pSZ}
&\approx
\frac{\tau^* \beta_{\rm p, \parallel}\,\beta^2_{\rm p, \perp}\,\expf{\mp 2 i \chi}}{20}\,
\Bigg[
\oOx{2}+\oOx{3}
\nonumber\\
&\qquad +\left(\oOx{2}+3\oOx{3}+\frac{29\oOx{4}}{21}+\frac{8\oOx{5}}{63}\right)\beta^2_{\rm p}
\Bigg]\,\nbb(x)
}
%---------------------------
for the related contributions. Here, we now also have a dependence on the line of sight component of the peculiar motion, $\beta_{\rm p, \parallel}=\beta_{\rm p}\,\mu_{\rm p}$.

Similarly, for $\ell=4$ we find
%---------------------------
\bealf{
\mathcal{P}^\pm_4(\vgh, \hat{\vb}_{\rm p})&=\sqrt{\frac{5}{18}}\,\left(1+\frac{7}{2}\,P_2(\mu_{\rm p})\right)(1-\mu_{\rm p}^2)\,\expf{\mp 2 i \chi}
\\ \nonumber
\Dbo{\pm, 0}{4 2 0}{0}(\nu, \beta_{\rm p})
&\approx \frac{\oOnu(1+\oOnu)^2(2+\oOnu)}{21\sqrt{15}}\,\beta^4_{\rm p}\Bigg[1
-
\left(\frac{2}{77}+\frac{25\oOnu}{77}-\frac{9\oOnu^2}{77}\right)\beta^2_{\rm p}
\Bigg],
}
%---------------------------
which then yields
%---------------------------
\bealf{
\label{eq:monopole_scattering_pSZ_sixth_4}
\Delta n^\pm\big|^{\ell=4}_{\rm pSZ}
&\approx
-\frac{\tau^* \beta^2_{\rm p, \perp}\,\expf{\mp 2 i \chi}}{210}\,\left(1+\frac{7}{2}\,P_2(\mu_{\rm p})\right)\beta^2_{\rm p}
\,\Bigg[
2\oOx{3}+\oOx{4}
\\ \nonumber
&\qquad +\left(4\oOx{3}+\frac{38\oOx{4}}{7}+\frac{128\oOx{5}}{77}+\frac{9\oOx{6}}{77}\right)\beta^2_{\rm p}
\Bigg]\,\nbb(x).
}
%---------------------------
Summing all contributions, we finally have 
%---------------------------
\bealf{
\label{eq:monopole_scattering_pSZ_tot}
&\Delta n^\pm\big|_{\rm pSZ}
\approx
-\frac{\tau^* \beta^2_{\rm p, \perp}\,\expf{\mp 2 i \chi}}{20}\,
\Bigg[
\oOx{2}+\left(\oOx{2}+\frac{4\oOx{3}}{3}+\frac{5\oOx{4}}{21}\right)\beta^2_{\rm p}
\\ \nonumber
&\qquad - \beta_{\rm p} P_1(\mu_{\rm p})\left(\oOx{2}+\oOx{3}\right)
+ \frac{\beta^2_{\rm p}}{3} P_2(\mu_{\rm p})\left(2\oOx{3}+\oOx{4}\right)\Bigg]\,\nbb(x)
}
%---------------------------
up to fourth order in $\beta_{\rm p}$. Although it is easy to generate higher order terms, we stop at this point, given that these terms become small.

%------------------------------------
\begin{figure}
    \centering
\includegraphics[width=\columnwidth]{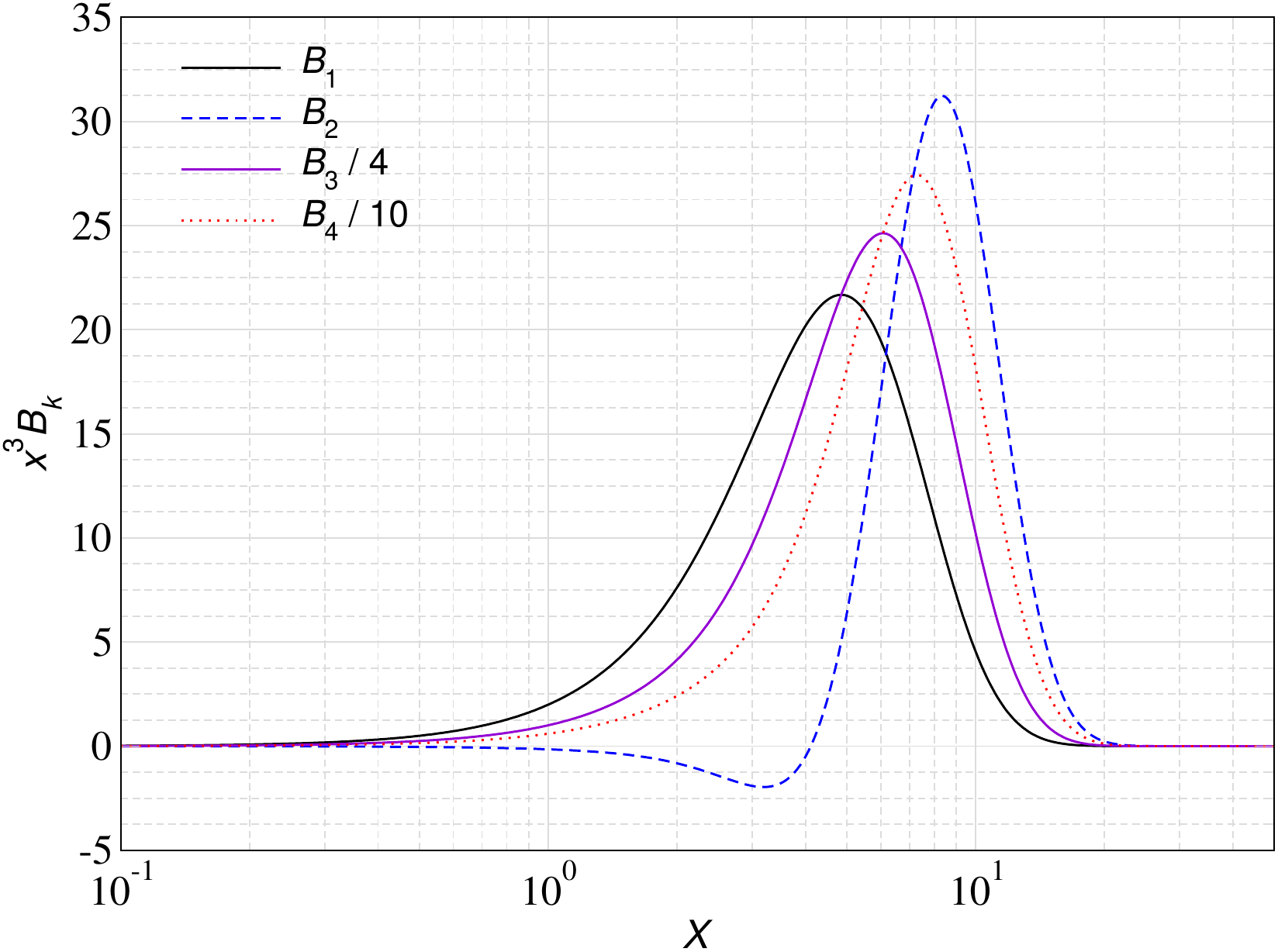}
\caption{Spectral functions $\mathcal{B}_k$. We normalized each of them to have a similar peak amplitude. The focus of the functions moves upwards as $k$ increases.}
    \label{fig:Bk}
\end{figure}
%------------------------------------
To illustrate the related spectra, let us define the functions
%---------------------------
\bsub
\bealf{
\label{eq:monopole_scattering_pSZ_tot_spectra}
\mathcal{B}_1(x)&=\oOx{2}\nbb(x)
\\
\mathcal{B}_2(x)&=\left(\oOx{2}+\frac{4\oOx{3}}{3}+\frac{5\oOx{4}}{21}\right)\nbb(x)
\\
\mathcal{B}_3(x)&=-\left(\oOx{2}+\oOx{3}\right)\nbb(x)
\\
\mathcal{B}_4(x)&=\left(2\oOx{3}+\oOx{4}\right)\nbb(x).
}
\esub
%---------------------------
The related spectra are show in Fig.~\ref{fig:Bk}. We can notice that the overall amplitude of the $\mathcal{B}_k$ increases with $k$. Similarly, the focus of the function moves to higher frequencies, relating to the repeated application of the energy shift operator, $\oOnu$. This leads to a differing relevance of the monopole, dipole and quadrupole dependence in terms of $\mu_{\rm p}$. However, since each of these is suppressed by powers of $\beta_{\rm p}$, the overall corrections beyond the leading order term are at the percent level or below. Note that the leading order octupole scattering terms $\propto \mathcal{B}_3$ provide sensitivity to the line of sight velocity component.

\subsubsection{Temperature corrections to the polarized SZ effect}
%---------------------------
It is quite easy to generate the relativistic temperature corrections to the polarized SZ effect using the expressions Eq.~\eqref{eq:monopole_scattering_pm_final}. Let us illustrate this for the terms $\beta_{\rm p}^2 \The$ and $\beta_{\rm p}^2 \The^2$, which were also considered earlier \citep[e.g.,][]{Challinor2000, Itoh2000Pol}. 

For this we need the operator $\Dbo{\pm, 0}{2 2 0}{0}(\nu, p)$ in orders of $p$, giving
%---------------------------
\bealf{
\Dbo{\pm, 0}{2 2 0}{0}(\nu, p)
&\approx \frac{\oOnu(1+\oOnu)}{3\sqrt{5}}\,p^2\Bigg[1
-
\left(\frac{10}{7}+\frac{3\oOnu}{7}-\frac{\oOnu^2}{7}\right)\,p^2
\\ \nonumber
&\qquad+\left(\frac{80}{49}+\frac{16\oOnu}{21}-\frac{76\oOnu^2}{441}-\frac{8\oOnu^3}{147}+\frac{4\oOnu^3}{441}\right)\,p^4
\Bigg]
\nonumber\\
&=
\frac{p^2}{3\sqrt{5}}\Bigg[\oOx{2}
+\left(8\oOx{3}+\oOx{4}\right)\frac{p^2}{7}\\ \nonumber
&\qquad\qquad+\left(\frac{360\oOx{4}}{7}+\frac{80\oOx{5}}{7}+\frac{4\oOx{6}}{7}\right) \frac{p^4}{63}
\Bigg],
}
%---------------------------
where in the last line we converted to powers of $\oOx{k}$. We also need 
%---------------------------
\bealf{
\label{eq:f_ell_approx}
f_{2}(\gamma, \gamma_{\rm p}) 
=f(\gamma)\,\frac{p^2 \beta^2_{\rm p}}{15\The^2} +\mathcal{O}(\beta^3_{\rm p}).
}
%---------------------------
Here, $f(\gamma)=\expf{-\gamma/\The}/[\The K_2(1/\The)]$ describes the thermal part of the distribution, determining the moments $\left<X(p)\right>=\int X(p)\,p^{2}f(\gamma)\id p$ of a quantity $X(p)$. Since $\left<p^4\right>\approx 15\The^2(1+6\The+15\The^2)$, $\left<p^6\right>\approx 105\The^3\left(1+\frac{21}{2}\The\right)$ and $\left<p^8\right>\approx 945 \The^4$, we then have 
%---------------------------
\bealf{
\label{eq:final_second_order}
\mathcal{\hat{S}}^\pm_{2}(\nu, \The, \beta_{\rm p})
&=-\frac{\sqrt{6}}{10}\,\sqrt{5}\,\left<\Dbo{\pm, 0}{2 2 0}{0}(\nu, p)\right>
\nonumber\\
&\approx -\frac{\beta_{\rm p}^2}{5\sqrt{6}}\,
\Bigg[\oOx{2}
+\left(6\oOx{2}+8\oOx{3}+\oOx{4}\right)\,\The\\ \nonumber
&\!\!+\left(15\oOx{2}+84\oOx{3}+\frac{867\oOx{4}}{14}+\frac{80\oOx{5}}{7}+\frac{4\oOx{6}}{7}\right)\,\The^2
\Bigg].
}
%---------------------------
Together with $\mathcal{P}^\pm_2(\vgh, \hat{\vb}_{\rm p})$ and $G(x)=x\expf{x}/(\expf{x}-1)^2$, this then yields
%---------------------------
\bsub
\bealf{
\label{eq:final_second_order_Dn}
&\Delta n^\pm\big|_{\rm pSZ}
\approx
-\frac{\tau^* \beta^2_{\rm p, \perp}\,\expf{\mp 2 i \chi}}{20}\,
\sum_{k=0}\The^k\,\mathcal{Y}_k(x)
\\
\mathcal{Y}_0&=\oOx{2}\,\nbb
\\
\mathcal{Y}_1&=\left(6\oOx{2}+8\oOx{3}+\oOx{4}\right)\,\nbb
\\
\mathcal{Y}_2&=\left(15\oOx{2}+84\oOx{3}+\frac{867\oOx{4}}{14}+\frac{80\oOx{5}}{7}+\frac{4\oOx{6}}{7}\right)\,\nbb
\\
\mathcal{Y}_3&=\Bigg(\frac{45\oOx{2}}{4}+399\oOx{3}+\frac{48873\oOx{4}}{56}+\frac{3480\oOx{5}}{7}
\nonumber\\
&\qquad+\frac{724\oOx{6}}{7}+\frac{60\oOx{7}}{7}+\frac{5\oOx{8}}{21}\Bigg)\,\nbb
\\
\mathcal{Y}_4&=\Bigg(-\frac{45\oOx{2}}{4}+945\oOx{3}+\frac{352215\oOx{4}}{56}+\frac{59100\oOx{5}}{7}
\\ \nonumber
&\!\!\!\!\!\!\!\!+\frac{12290\oOx{6}}{3}+\frac{18610\oOx{7}}{21}+\frac{3865 \oOx{8}}{42}+\frac{40\oOx{9}}{9}+\frac{5 \oOx{10}}{63}\Bigg)\,\nbb
}
\esub
%---------------------------
where $\mathcal{Y}_0, \mathcal{Y}_1$ and $\mathcal{Y}_2$ directly follow from above and agree with previous results \citep[e.g.,][]{Itoh2000Pol}. The other terms we give for demonstration of the method. Higher order terms can be easily obtained using {\tt Mathematica}.\footnote{A {\tt Mathematica} file is available here:  \url{www.chluba.de/Mathematica}}

%------------------------------------
\begin{figure}
    \centering
\includegraphics[width=\columnwidth]{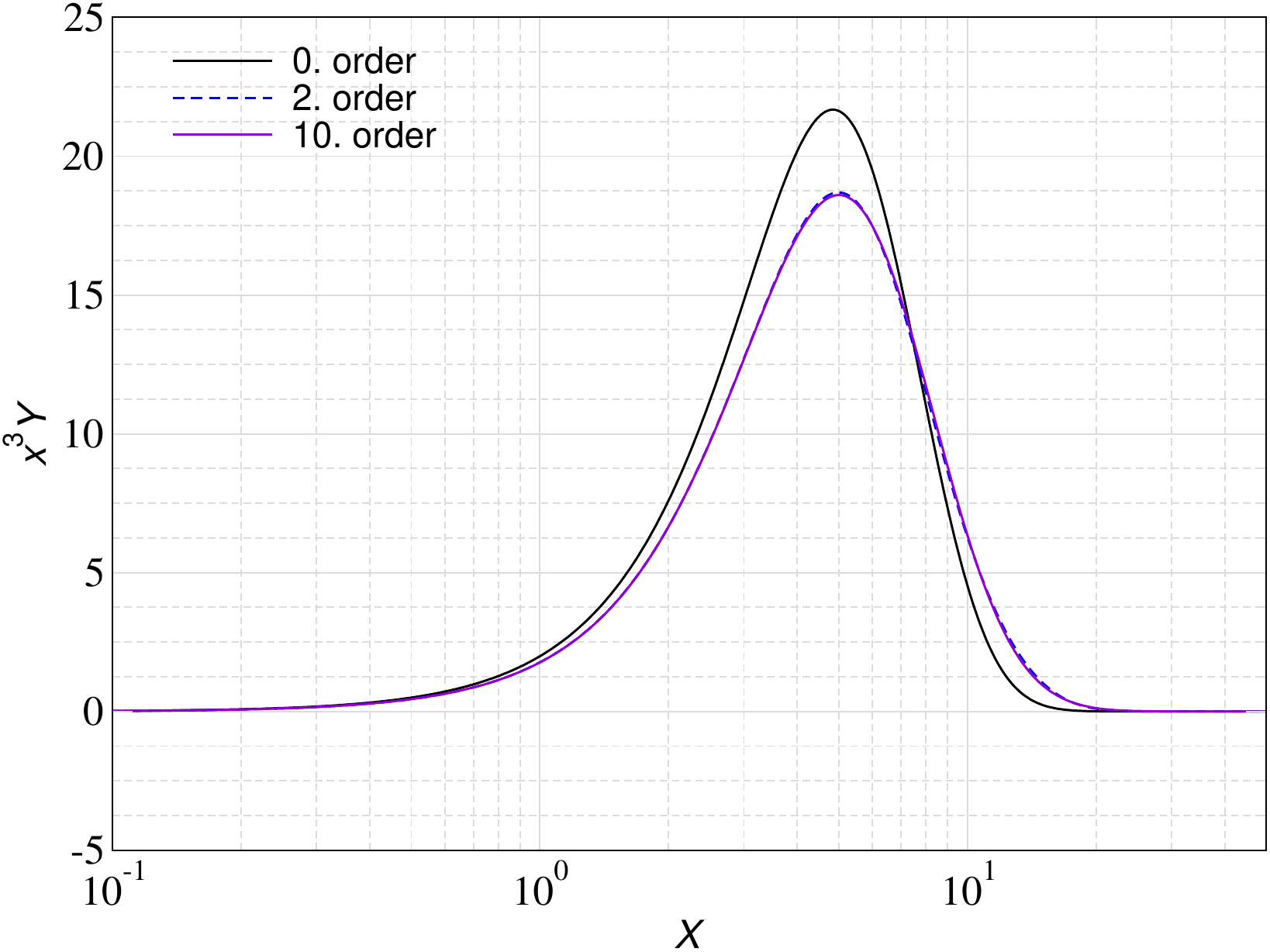}
\caption{Spectral function $\mathcal{Y}=\sum_{k=0}^{k_{\rm max}} \The^k \mathcal{Y}_k$ up to varying orders and for $\The=0.02$ (i.e, $\simeq 10\,{\rm keV}$).}
    \label{fig:pSZ}
\end{figure}
%------------------------------------
In Fig.~\ref{fig:pSZ} we illustrate the spectrum of the generated polarized radiation as a function of $x$. We assumed $\The=0.02$ (i.e, $\simeq 10\,{\rm keV}$) for the computation and varied the included temperature order. The leading order term does not capture the spectrum well, while corrections up to second order in temperature are already quite accurate. Like for the normal thermal SZ effect, precision of the expressions (even up to 10. order in the $\The$) becomes worse at high frequencies ($x>5-10$) due to the asymmptotic convergence of the series \citep[e.g.,][]{Itoh98, Chluba2012SZpack}.

\subsection{Scattering of primary CMB anisotropies}
\label{sec:aniso_scattering}
%---------------------------
We now consider the scattering of the primary CMB anisotropies by an isotropic thermal electron distribution, such that Eq.~\eqref{eq:S_final_iso} is applicable. The scattering of the intensity monopole and dipole does not lead to any polarization terms in the considered case. However, starting with a temperature quadrupole, $n^0_{2m}=\Theta_{2m} G(x)$, at first order in the electron temperature we find 
%---------------------------
\bsub
\bealf{
\label{eq:ndot_p0_Qsc}
\Delta n^0_{2 m}(\nu)
&\approx \tau \left\{-\frac{9}{10}
+\frac{\The}{10}(\oDnu-6)\right\} 
\,n_{2 m}^0(\nu)
\\
\Delta n^\pm_{2 m}(\nu)
&\approx -\frac{\sqrt{6}}{10}\tau \left\{1
+\The(\oDnu-4)
\right\} 
\,n_{2 m}^0(\nu)
}
\esub
%---------------------------
after one scattering. We observe the sourcing of $n^\pm$ at leading order, with some temperature correction. We also have $$\oDnu G(x)=\oOnu \oDnu \nbb = -[4 \oOx{} + 6 \oOx{2}+\oOx{3}]\nbb= Y_1(x),$$
a function that appears in the generation of spectral distortion anisotropies \citep{chluba_spectro-spatial_2023-I} and causes a small change in the spectrum here. Comparing to \citet{Challinor2000}, Eq.~(32), we confirm that $\mathcal{Q}_1(x)=(\oDnu-4)G(x)$ corresponds to the first temperature correction term given there. Similarly, taking the second order temperature terms from ${_{\pm 0}}\hat{\mathcal{S}}_{2}$ in Eq.~\eqref{eq:S_explicit_s0}, we find 
%---------------------------
\bealf{
\label{eq:ndot_p0_Qsc_The2}
\Delta n^{\pm, \The^2}_{2 m}(\nu)
&\approx -\frac{\sqrt{6}}{10}\tau \The^2 \left\{\frac{90}{7}-\frac{11\oDnu}{2}+\frac{4\oDnu^2}{7}\right\} 
\,n_{2 m}^0(\nu),
}
%---------------------------
which with $n^0_{2m}=\Theta_{2m} G(x)$ gives the spectrum 
%---------------------------
\bealf{
\label{eq:ndot_p0_Qsc_The2_F}
\mathcal{Q}_2(x)
&= \frac{90}{7}\,G(x)-\frac{11}{2} Y_1(x)-\frac{12}{7}Y_2(x)+\frac{4}{7} Y_3(x),
}
%---------------------------
where $Y_k(x)=\oOnu^k Y(x)$ and $\oDnu Y_k=(\oOnu^2-3\oOnu)\,Y_k=Y_{k+2}-3Y_{k+1}$.This reproduces Eq.~(33) of \citet{Challinor2000}. For convenience we give all terms up to 5th order in the temperature, such that
%---------------------------
\bealf{
\label{eq:ndot_p0_Qsc_The5}
\Delta n^{\pm}_{2 m}(\nu)
&\approx -\frac{\sqrt{6}}{10} \tau\left[G(x)+\sum_{k=1}\The^k \mathcal{Q}_k(x)\right]
\,\Theta_{2 m}^0
\nonumber\\
\mathcal{Q}_3(x)&=-\frac{585}{14}G+\frac{3931}{168}Y_1+\frac{88}{7}Y_2-\frac{103}{21}Y_3+\frac{5}{21}Y_4
\nonumber\\
\mathcal{Q}_4(x)&=\frac{2105}{14}G-\frac{49345}{504}Y_1-\frac{408}{7}Y_2+\frac{415}{14}Y_3-\frac{5}{2}Y_4+\frac{5}{63}Y_5
\nonumber\\
\mathcal{Q}_5(x)&=-\frac{138225}{224}G+\frac{1176275}{2688}Y_1+\frac{5341}{15}Y_2-\frac{83131}{504}Y_3
\nonumber\\
&\qquad +\frac{731}{40}Y_4-\frac{107}{105}Y_5+\frac{1}{45}Y_6.
}
%---------------------------
Similarly, for polarization from the scattering of the octupole we find
%---------------------------
\bealf{
\label{eq:ndot_p0_Qsc_The5}
\Delta n^{\pm}_{2 m}(\nu)
&\approx \frac{\sqrt{6}}{14\sqrt{5}} \tau \left[\sum_{k=1}\The^k \mathcal{O}_k(x)\right]
\,\Theta_{3 m}^0
\nonumber\\
\mathcal{O}_1(x)&=\mathcal{F}_1(x)
\nonumber\\
\mathcal{O}_2(x)&=30G-\frac{23}{2}Y_1-3Y_2+Y_3
\nonumber\\
\mathcal{O}_3(x)&=-\frac{335}{2}G+\frac{5767}{72}Y_1+\frac{106}{3}Y_2-\frac{121}{9}Y_3+\frac{5}{9}Y_4
\nonumber\\
\mathcal{O}_4(x)&=\frac{1775}{2}G-\frac{34951}{72}Y_1-277 Y_2+\frac{689}{6}Y_3-\frac{49}{6}Y_4+\frac{2}{9}Y_5
\nonumber\\
\mathcal{O}_5(x)&=-\frac{1704525}{352}G+\frac{12207463}{4224}Y_1+\frac{253609}{132}Y_2-\frac{668957}{792}Y_3
\nonumber\\
&\qquad +\frac{20755}{264}Y_4-\frac{41}{11}Y_5+\frac{7}{99}Y_6.
}
%---------------------------
which as expected vanishes at zero temperature.

If instead we start with a polarization quadrupole, $n^\pm_{2m}=\Theta^\pm_{2m} G(x)$, at first order in the electron temperature we find 
%---------------------------
\bealf{
\label{eq:ndot_p0_Qsc}
\Delta n^0_{2 m}(\nu)
&\approx -\frac{\sqrt{6}}{10}\tau \left\{1+ \The(\oDnu-4)\right\} n^Q_{2m}
\,
\\ \nonumber 
\Delta n^Q_{2 m}(\nu)
&\approx -\frac{2\tau}{5} \left\{1+\frac{3}{2}\The\left(2 - \oDnu\right)\right\}n^Q_{2m}
\\ \nonumber 
\Delta n^U_{2 m}(\nu)
&\approx -\tau \left\{1+\frac{2\The}{5}\left(2 + \oDnu\right)\right\}n^U_{2m},
}
%---------------------------
where we transformed to $n^Q_{2m}=\frac{n_{2 m}^++n_{2 m}^-}{2}$ and $n^U_{2m}=\frac{n_{2 m}^+-n_{2 m}^-}{2 i}$. Again we can see that temperature corrections cause evolution of the photon spectrum through the diffusion operator. It is worth noting that the distortions in the spectra of $Q$ and $U$ differ slightly.

It is rather easy to include higher order diffusion corrections and even add kinematic corrections. With the provided {\tt Mathematica} notebook this can be done as required. The boost operator approach is therefore essentially a complete analytic solution of the polarized scattering problem in the Thomson limit.

\section{Conclusions}
\label{sec:Conc}
We have used the boost operator approach of \citet{ChlubaBO25} to derive exact expressions for the polarized SZ effect from both intrinsic intensity and polarization isotropies and those induced by bulk motion of the electrons, with all thermal corrections. 
Equation~\eqref{eq:final_general_expressions_S_final_kin} gives the most general result for the evolution of polarized radiation from a moving electron cloud in the Thomson limit in a compact expression.
For the special case of isotropic electron momenta we computed the direction-averaged scattering operator in Eq.~\eqref{eq:S_final_iso}, finding that in this case there was no coupling between different $(\ell, m)$ states but non-trivial mixing of intensity and polarization states as well as rotation of the polarization direction; these physical effects were further illustrated by expanding the expression in orders of electron momentum. These operator expressions can in principle be applied to any photon field for which recoil and stimulated effects remain negligble.

In Sec.~\ref{sec:cmb_apps} we worked the examples of pure kinematic polarized SZ with temperature corrections when scattering off the CMB monopole, as well as the CMB quadrupole and octupole for the case of isotropic electrons, reproducing all previous results in these cases.
These examples can be easily reproduced and extended to other cases using the {\tt Mathematica} notebook that has been provided.
This is a direct illustration of the use of the boost operator to solve the polarized relativistic scattering problem exactly, and is generally extendable to additional cases such as multiple scattering, non-black body spectra, and non-negligible electron recoil.

We also note that in this context we find that all the required spectra can be related to boosts of the standard $y$-distortion. The basis spectra can be written as $Y_k=\oOnu^k Y(x)$, functions that also appear in recently applications to anisotropic spectral distortions \citep{chluba_spectro-spatial_2023-I, chluba_spectro-spatial_2023-II, kite_spectro-spatial_2023-III}. This is because all spectra are generated by superpositions of boosts which are fully spanned by the $Y_k$ function. We can therefore also treat the evolution of anisotropic spectral distortions in the presence of relativistic electron populations using the same basis, a problem that we shall investigate in the future.

Finally, here we omitted recoil and stimulated recoil terms. These can in principle be included as part of the scattering operator in the electron rest frame. Including only the first order recoil term, one should be able to derive the Kompaneets equation equivalent for polarization states. The proposed matrix decomposition given in Sec.~\ref{sec:scat_iso_matrix} should provide a very intuitive way to collect the required terms, another problem we leave to future work.

%----------------------------------

\small 

\vspace{2mm}

\noindent
{\it Data Availability Statement}: {\tt Mathematica} files to reproduce some of the key results are available at \url{www.chluba.de/Mathematica}.

\vspace{3mm}

\noindent
{\it Acknowledgments}: The authors acknowledge support from the SO:UK project, grant reference ST/X006344/1. We also acknowledge support from the MMT exchange program, which stimulated the ideas for this paper as part of the MMT workshop in Manchester, May 2025.

\bibliographystyle{mn2e}
\bibliography{Lit-2025,polsz}

\onecolumn

\begin{appendix}

\section{Computation of the spin harmonics}
\label{app:spin_harmonics}
%---------------------------
To generate the functions ${_{\pm 2}}Y_{\ell m}(\vgh)$, we can use \citep{Goldberg1967, Hu1997}
%------------------------------
\bealf{
{_{\pm 2}}Y_{\ell m}(\vgh)&=
\sqrt{\frac{(\ell-2)!}{(\ell+2)!}}
\left[
\partial^2_\theta-\cot \theta \, \partial_\theta\pm 
\frac{2 i}{\sin \theta}(\partial_\theta-\cot \theta)\,\partial_\phi-\frac{1}{\sin^2 \theta}\partial^2_\phi
\right] Y_{\ell m}(\vgh).
}
%------------------------------
Defining $\mu=\cos \theta$, this can be rewritten as
%------------------------------
\bealf{
{_{\pm 2}}Y_{\ell m}(\vgh)&=
\sqrt{\frac{(\ell-2)!}{(\ell+2)!}}
\left[(1-\mu^2) \partial^2_\mu \mp 
2 i\left\{\partial_\mu+\frac{\mu}{\sin^2 \theta}\right\}\,\partial_\phi-\frac{1}{\sin^2 \theta}\partial^2_\phi
\right] Y_{\ell m}(\vgh)
\nonumber\\
&=\sqrt{\frac{2\ell +1}{4\pi}\frac{(\ell-m)!}{(\ell+m)!}\frac{(\ell-2)!}{(\ell+2)!}}\,\expf{i m \phi}
\left[(1-\mu^2) \partial^2_\mu  \pm 
2 m \left\{\partial_\mu+\frac{\mu}{\sin^2 \theta}\right\}+\frac{m^2}{\sin^2 \theta}
\right]P_{\ell}^m(\mu)
\nonumber\\
&=\sqrt{\frac{2\ell +1}{4\pi}\frac{(\ell-m)!}{(\ell+m)!}\frac{(\ell-2)!}{(\ell+2)!}}\,\expf{i m \phi}
\left[2 (\mu\pm m)\, \partial_\mu  
\pm
\frac{2 m (\mu\pm m) }{1-\mu^2}-\ell(\ell+1)\right] P_{\ell}^m(\mu),
}
%------------------------------
where in the last step we used the differential equation for the associated Legendre polynomials.
Written in this way, we can directly see that for $m=0$ we have 
%------------------------------
\bealf{
{_{\pm 2}}Y_{\ell 0}(\vgh)&=
\sqrt{\frac{2\ell +1}{4\pi}\frac{(\ell-2)!}{(\ell+2)!}}\,
(1-\mu^2) \,\partial^2_\mu \,P_{\ell}(\mu) 
=\sqrt{\frac{2\ell +1}{4\pi}\frac{(\ell-2)!}{(\ell+2)!}}\left[2\mu\partial_\mu -\ell(\ell+1)\right]P_{\ell}(\mu)
=\sqrt{\frac{2\ell +1}{4\pi}\frac{(\ell-2)!}{(\ell+2)!}}\left[2\partial_\mu P_{\ell+1}(\mu)-(\ell+2)(\ell+1)P_{\ell}(\mu)\right]
}
%------------------------------
in terms of the Legendre polynomials. The remaining derivative can be rewritten as
%------------------------------
\bealf{
\partial_\mu P_{\ell+1}(\mu)
&=(2\ell +1)P_{\ell}(\mu)+(2[\ell-2] +1)P_{\ell-2}(\mu)+(2[\ell-4] +1)P_{\ell-2}(\mu)+\ldots
=
\sum_{j\,\text{even/odd}}^{\ell-2} (2j+1)P_j(\mu),
}
%------------------------------
giving the expression for $m=0$ spin harmonics
%------------------------------
\bealf{
{_{\pm 2}}Y_{\ell 0}(\vgh)&=
\sqrt{\frac{2\ell +1}{4\pi}\frac{(\ell-2)!}{(\ell+2)!}}\left[
-(\ell+2)(\ell+1)P_{\ell}
+\sum_{j\,\text{even/odd}}^{\ell-2} 2(2j+1)P_j(\mu)(\mu)\right].
}
%------------------------------
Here, the sum over $j$ goes only over even/odd numbers for even/odd $\ell$.

\vspace{-3mm}
\subsection{Explicit expressions and relation to Wigner D-matrix}
\label{app:sYlm_explicit}
%---------------------------
One can show that the spin-weighted harmonic functions can be explicitly given as \citep{Goldberg1967}
%---------------------------
\bealf{
\label{eq:sYlm_def_gen}
{_{s}}Y_{\ell m}(\theta, \phi)&=(-1)^{\ell+m-s}\,\sqrt{\frac{2\ell+1}{4\pi}\,\frac{(\ell+m)!(\ell-m)!}{(\ell+s)!(\ell-s)!}}\,
\sin^{2\ell}\left(\frac{\theta}{2}\right)\,\expf{im\phi}\,
\sum_{r=0}^{\ell-s}(-1)^r
\,\binom{\ell-s}{r}
\,\binom{\ell+s}{r+s-m}
\,\cot^{2r+s-m}\left(\frac{\theta}{2}\right),
}
%---------------------------
which here includes a factor of $(-1)^m$ to match the standard spherical harmonic definitions. We can compare this to the Wigner D-matrix, which is defined by the rotation of the coordinate system by the Euler angles \citep[see][Sec. 4.3.1, Eq.~(5)]{Varshalovich1988}
%---------------------------
\bealf{
D_{m m'}^\ell(\alpha, \beta, \gamma)&=\bra{\ell m} \,\hat{R}(\alpha, \beta, \gamma) \,\ket{\ell m'}=\bra{\ell m}\, \expf{-i \alpha \hat{J}_z}\, \expf{-i \beta \hat{J}_y}\, \expf{-i \gamma \hat{J}_z} \,\ket{\ell m'}=\expf{-i m \alpha} \, d^\ell_{m m'}(\beta)\,\expf{-i m' \gamma}
\\
d^\ell_{m m'}(\beta)&=
\sqrt{(\ell+m)!(\ell-m)!(\ell+m')!(\ell-m')!}\,
\sum_{t}
\frac{(-1)^{m-m'+t} \cos^{2\ell+m'-m-2t}\left(\frac{\beta}{2}\right)\sin^{m-m'+2t}\left(\frac{\beta}{2}\right)}{(\ell+m'-t)! (\ell-m-t)! t! (m-m'+t)!},
}
%---------------------------
where the sum over $t$ needs to respect that the arguments of the factorials remain non-negative. 
From these expressions we can immediately read of $D_{m m'}^{\ell*}(\alpha, \beta, \gamma)=D_{m m'}^\ell(-\alpha, \beta, -\gamma)=(-1)^{m-m'}\,D_{m m'}^\ell(-\alpha, -\beta, -\gamma)=D_{m' m}^\ell(-\gamma, -\beta, -\alpha)=[\hat{R}^{-1}(\alpha, \beta, \gamma)]^\ell_{m' m}$, the latter signifying the inverse rotation of the system. We also have  $d^\ell_{m m'}(-\beta)=d^\ell_{m' m}(\beta)=(-1)^{m+m'} d^\ell_{m m'}(\beta)$ and $d^\ell_{m m'}(\beta)=d^\ell_{-m', -m}(\beta)$.
By defining $r=\ell+m'-t$ and pulling out a factor of $\sin^{2\ell}(\beta/2)$, we can also \citep[see][Sec. 4.3.1, Eq.~(3)]{Varshalovich1988}
%---------------------------
\bealf{
d^\ell_{m m'}(\beta)
&=
\sqrt{(\ell+m)!(\ell-m)!(\ell+m')!(\ell-m')!}\,
\,
\sin^{2\ell}\left(\frac{\beta}{2}\right)\,
\sum_{r}
\frac{(-1)^{\ell + m - r} \cot^{2r-m'-m}\left(\frac{\beta}{2}\right)}{r! (r-m'-m)! (\ell+m'-r)! (\ell+m-r)!}
\nonumber \\[-0.5mm]
&=
(-1)^{\ell + m}\,\sqrt{\frac{(\ell+m)!(\ell-m)!}{(\ell+m')!(\ell-m')!}}\,
\,
\sin^{2\ell}\left(\frac{\beta}{2}\right)\,
\sum_{r}
\frac{(-1)^{r} (\ell+m')!(\ell-m')! \cot^{2r-m'-m}\left(\frac{\beta}{2}\right)}{r! (r-m'-m)! (\ell+m'-r)! (\ell+m-r)!}
\nonumber \\[-0.5mm]
&=
(-1)^{\ell + m}\,\sqrt{\frac{(\ell+m)!(\ell-m)!}{(\ell+m')!(\ell-m')!}}\,
\,
\sin^{2\ell}\left(\frac{\beta}{2}\right)\,
\sum_{r}
(-1)^{r}\,\binom{\ell+m'}{r}\,\binom{\ell-m'}{r-m'-m}\,
\cot^{2r-m'-m}\left(\frac{\beta}{2}\right).
}
%---------------------------
which agrees with \citet{Goldberg1967} when $\beta \rightarrow -\beta$ and after correcting the Condon-Shortley phase. By comparing this with Eq.~\eqref{eq:sYlm_def_gen}, we can see that with $s=-m'$, we then have
%---------------------------
\bealf{
\label{eq:D_using_Ylm}
D_{m m'}^\ell(\alpha, \beta, \gamma)&\equiv
(-1)^{m'}\sqrt{\frac{4\pi}{2\ell+1}}\,{_{-m'}}Y^*_{\ell m}(\beta, \alpha)\,\expf{-im'\gamma}
\equiv 
(-1)^{m}\sqrt{\frac{4\pi}{2\ell+1}}\,{_{m'}}Y_{\ell ,-m}(\beta, \alpha)\,\expf{-im'\gamma}
}
%---------------------------
for our definition of the rotations. Note that this expression is consistent with the one given in \citet{Goldberg1967} when recognizing that the Wigner D-symbols are there written for the inverse rotation, $\hat{R}^{\dagger}(\alpha, \beta, \gamma)= \expf{i \gamma \hat{J}_z}\, \expf{i \beta \hat{J}_y}\, \expf{i \alpha \hat{J}_z}=\hat{R}(-\gamma, -\beta, -\alpha)$. Adding the overall Condon-Shortley phase, this implies
%---------------------------
\bealf{
\label{eq:D_Goldberg}
(-1)^{m'-m}\,D_{m m'}^{\ell, \rm G}(\alpha, \beta, \gamma)
%=\left< \ell m \,\big|\, \hat{R}^\dagger(\alpha, \beta, \gamma) \,\big|\, \ell m'\right>\bigg|_{\rm G}
=\left< \ell m \,\big|\, \hat{R}(-\gamma, -\beta, -\alpha) \,\big|\, \ell m'\right>=D_{m m'}^{\ell}(-\gamma, -\beta, -\alpha)=\expf{i m \gamma} \, d^\ell_{m m'}(-\beta)\,\expf{i m' \alpha}=\expf{i m' \alpha}\, d^\ell_{m' m}(\beta)\,\expf{i m \gamma}\equiv 
D_{m' m}^{\ell *}(\alpha, \beta, \gamma).
}
%---------------------------
Explicitly, we also have \citep{Goldberg1967, Hu1997}
%---------------------------
\bealf{
\label{eq:D_Goldberg_Ylm}
D_{s m}^{\ell, \rm G}(\alpha, \beta, \gamma)
&=
\sqrt{\frac{4\pi}{2\ell+1}}\,{_{-s}}Y^{\rm G}_{\ell m}(\beta, \alpha)\,\expf{i s\gamma}
=
(-1)^m\,\sqrt{\frac{4\pi}{2\ell+1}}\,{_{-s}}Y_{\ell m}(\beta, \alpha)\,\expf{i s\gamma}
\equiv (-1)^{m-s}\,D_{m s}^{\ell *}(\alpha, \beta, \gamma).
}
%---------------------------
For our derivations, we use the identity Eq.~\eqref{eq:D_using_Ylm}. We also note that even if \citet{Dai2014} used the convention of  \citet{Goldberg1967} in their derivations, the results for the kernel should remain unchanged. 

\subsection{Transformation of spin harmonics and harmonic coefficients}
\label{app:trans_law_Ylm_alm}
%---------------------------
With the definitions above, we can write the transformed spin harmonics as \citep{Varshalovich1988}
%---------------------------
\bealf{
\label{eq:trans_Ylm}
\left|\, s \ell m\right>'&
\equiv {_{s}}Y_{\ell m}(\theta', \phi')
\equiv\hat{R}(\alpha, \beta, \gamma)\,\left|\, s \ell m\right>=
\sum_{\ell' m'} \left|\, s \ell' m'\right>\left<\, s \ell' m'\,\right|\,\hat{R}(\alpha, \beta, \gamma)\,\left|\, s \ell m\right>
=\sum_{m'}\,{_{s}}Y_{\ell m'}(\theta, \phi)\,D_{m' m}^\ell(\alpha, \beta, \gamma)
}
%---------------------------
This then means that the harmonic coefficients of a function $X(\theta, \phi)=\sum_{\ell m} {_{s}}X_{\ell m}\,{_{s}}Y_{\ell m}(\theta, \phi)=\sum_{\ell m} {_{s}}X_{\ell m} \left| s \ell m \right>$ transform as
%---------------------------
\bealf{
\label{eq:trans_alm}
{_{s}}X'_{\ell m}&=\left<\, s \ell m\,\right|'
\,\sum_{\ell' m'} {_{s}}X_{\ell m'} \left| s \ell' m' \right>
=\sum_{m'} {_{s}}X_{\ell m'} 
\left<\, s \ell m\,\right|\,\hat{R}^\dagger(\alpha, \beta, \gamma)\,\left|\, s \ell m'\right>
=\sum_{m' m''} D_{m'' m}^{\ell *}(\alpha, \beta, \gamma) \,{_{s}}X_{\ell m'}\,
\int {_{s}}Y^*_{\ell m''} {_{s}}Y_{\ell m'}\id^2\Omega
=\sum_{m'} D_{m' m}^{\ell *}(\alpha, \beta, \gamma) \,{_{s}}X_{\ell m'}.
}
%---------------------------
This expression directly implies
%---------------------------
\bealf{
\label{eq:trans_alm}
X'(\theta', \phi')&=\sum_{\ell m} {_{s}}X'_{\ell m}\,{_{s}}Y_{\ell m}(\theta', \phi')=
\sum_{\ell m m'} D_{m' m}^{\ell *}(\alpha, \beta, \gamma)\,{_{s}}X_{\ell m'}\,{_{s}}Y_{\ell m}(\theta', \phi')=
\sum_{\ell m m' m''} D_{m' m}^{\ell *}(\alpha, \beta, \gamma)\,{_{s}}X_{\ell m'}\,D_{m'' m}^{\ell}(\alpha, \beta, \gamma)\,{_{s}}Y_{\ell m''}(\theta, \phi)=X(\theta, \phi)
}
%---------------------------
where in the last step we used $\sum_m D_{m' m}^{\ell *}(\alpha, \beta, \gamma)\,D_{m'' m}^{\ell}(\alpha, \beta, \gamma)=\delta_{m' m''}$, which follows from $\hat{R}$ being unitary.

\subsection{Generalized addition theorem for spin harmonics}
\label{app:addition_sYlm}
%---------------------------
Similar to the addition theorem for $s=0$ spherical harmonics
%---------------------------
\bealf{
\label{eq:addition_Ylm}
%P_\ell(\vgh\cdot\vghp)=\frac{4\pi}{2\ell+1}\,\sum_m Y_{\ell m}(\vgh)\,Y^*_{\ell m}(\vghp)=\frac{4\pi}{2\ell+1}\,\sum_m Y^*_{\ell m}(\vgh)\,Y_{\ell m}(\vghp)
\sum_m Y_{\ell m}(\vgh)\,Y^*_{\ell m}(\vghp)=\sum_m Y^*_{\ell m}(\vgh)\,Y_{\ell m}(\vghp)\equiv 
\frac{2\ell+1}{4\pi}P_\ell(\vgh\cdot\vghp)
}
%---------------------------
one can find a generalized form for spin harmonics \citep{Hu1997}:
%---------------------------
\bealf{
\label{eq:addition_sYlm}
\sum_m {_{s_1}}Y^{*}_{\ell m}(\vgh)
\,{_{s_2}}Y_{\ell m}(\vghp)
=\sum_m {_{s_1}}Y^{\rm G *}_{\ell m}(\vgh)
\,{_{s_2}}Y^{\rm G}_{\ell m}(\vghp)
%=(-1)^{s_1+s_2} \sum_m {_{s_1}}Y^{\rm G}_{\ell m}(\vgh)\,{_{s_2}}Y^{\rm G *}_{\ell m}(\vghp)
=\sqrt{\frac{2\ell+1}{4\pi}}
{_{s_2}}Y^{\rm G}_{\ell, -s_1}(\theta, \alpha)
\,\expf{-i s_2 \chi}
=(-1)^{s_1}\sqrt{\frac{2\ell+1}{4\pi}}
{_{s_2}}Y_{\ell, -s_1}(\theta, \alpha)
\,\expf{-i s_2 \chi}.
}
%---------------------------
Here, $\cos\theta=\vgh\cdot\vghp$, $\alpha$ is the angle that $\vgh-\vghp$ makes with $\hat{\vek{e}}_{\theta}(\vghp)$, and $\chi$ is the angle that $\vgh-\vghp$ makes with $\hat{\vek{e}}_{\theta}(\vgh)$ \citep[cf. Eq.~(7) and Fig.~1 of][]{Hu1997}. The overall phase factor is because \citet{Hu1997} follow the conventions of \citet{Goldberg1967} with respect to the Condon-Shortley phase.

\section{Computations of the kernel elements}
\label{app:Kernel_computation}
%---------------------------
For our problem we require the aberration kernel elements $\Kk{d}{0}{m}{\ell \ell'}(-\beta)$ and $\Kk{d}{\pm 2}{m}{\ell \ell'}(-\beta)$. These can be computed from the initial conditions $\Kk{d}{0}{m}{mm}(-\beta)$ and $\Kk{d}{\pm 2}{m}{mm}(-\beta)$ using the properties of the kernel and $\ell$-raising operations \citep{Chluba2011ab, Dai2014}. 

\subsection{Raising operations for kernel elements}
\label{app:Kernel_raise}
%---------------------------
We first generalize the recursion relation given in the Appendix of \citet{ChlubaBO25b} to general $m$ and $s$. Using relations for $_sY_{\ell m}$ given in \citet{Dai2014} Appendix~C, we can write
%---------------------------
\begin{equation}
{_s}Y_{\ell m}(\vghp) = \frac{1}{{_s}C_\ell^m}\left[\left(\mu' + \frac{sm}{l(l-1)}\right) {_s}Y_{l-1 m}(\vghp) - {_s}C_{\ell-m}^m {_s}Y_{\ell-2 m}(\vghp) \right].   
\end{equation}
%---------------------------
Realizing that $\mu' = (D-\gamma)/p$ we then obtain\footnote{Note the '$-$' sign flip in the $s$-term from the definition of the aberration kernel.} 
%---------------------------
\begin{equation}
\label{eq:raise_spin2}
\Kk{d}{s}{m}{\ell \ell'}(-\beta) =- \frac{1}{{_s}C_\ell^m}\,\frac{\gamma \,\Kk{d}{s}{m}{\ell-1 \ell'}(-\beta) - \Kk{d-1}{s}{m}{\ell-1 \ell'}(-\beta)}{p}
-
\frac{1}{{_s}C_\ell^m}\frac{sm}{l(l-1)}\, \Kk{d}{s}{m}{\ell-1 \ell'}(-\beta) - \frac{{_s}C_{\ell-1}^m}{{_s}C_\ell^m} \,\Kk{d}{s}{m}{\ell-2 \ell'}(-\beta)
\end{equation}
%---------------------------
where ${_s}C_\ell^m = \sqrt{(\ell^2-m^2)(\ell^2-s^2)/(4\ell^2-1)}/\ell$. This expression also follows from Eq.~(7) of \citet{Dai2014}.

\subsection{Generating the kernel elements for $s=0$}
\label{app:Kernel_s0}
%---------------------------
To compute all aberration kernel elements $\Kk{d}{0}{m}{\ell \ell'}(-\beta)$, we can use the following procedure \citep{Chluba2011ab, Dai2014}: starting with the elements 
%---------------------------
\begin{align}
\Kk{d}{0}{m}{mm}(-\beta)
&= \int  \frac{Y{_{mm}^*}(\vghp)\, Y_{mm}(\vgh[\vghp])}{[\gamma(1+\beta \mu')]^d} \id\vghp \equiv 
\frac{[(2m-1)!!]^2\,(2m+1)}{2 (2m)!}
\int  \frac{(1-{\mu'}^2)^{m/2}(1-{\mu}^2[\mu'])^{m/2}}{[\gamma(1+\beta \mu')]^d} \id\mu'
= \frac{1}{\gamma^{m+d}}\,{}_2 F_1\left( \frac{m+d}{2}, \frac{m+d+1}{2}, \frac{3}{2}+m, \frac{p^2}{\gamma^2}\right),
\end{align}
%---------------------------
where ${}_2 F_1(a, b, c, z)$ is the hypergeometric function. Here, we also used $P^m_m(x)=(-1)^m\,(2m-1)!!\,(1-x^2)^{m/2}$ and $\mu[\mu']=(\mu'+\beta)/(1+\beta \mu')$. We can initialize the sequence $\Kk{d}{0}{m}{mm}, \Kk{d}{0}{m}{m+1, m}, \Kk{d}{0}{m}{m+2, m}, \ldots $ to any $l_{\rm max}$. Using $\Kk{d}{0}{m}{\ell\ell'}(-\beta)=\Kk{d}{0}{m}{\ell'\ell}(\beta)$, we then can obtain all elements $\Kk{d}{0}{m}{mm}, \Kk{d}{0}{m}{m, m+1}, \Kk{d}{0}{m}{m, m+2}, \ldots$ which then allows us to start the next sequence $\Kk{d}{0}{m}{m, m+1}, \Kk{d}{0}{m}{m+1, m+1}, \Kk{d}{0}{m}{m+2, m+1}, \ldots $, a process that we can repeat until satisfied. Since $\Kk{d}{s}{m}{\ell\ell'}(-\beta)=\Kk{d}{-s}{-m}{\ell\ell'}(-\beta)$ for $s=0$ implies $\Kk{d}{0}{m}{\ell\ell'}(-\beta)=\Kk{d}{0}{-m}{\ell\ell'}(-\beta)$, we then automatically have all kernels for $m<0$ too.  This is implemented in the provided {\tt Mathematica} notebook and is significantly faster than performing the explicit computations.

\subsection{Generating the kernel elements for $s=\pm 2$}
\label{app:Kernel_spm2}
%---------------------------
For $s=\pm 2$ we actually only require the kernel elements $\Kk{d}{\pm 2}{m}{\ell 2}(-\beta)$ and $\Kk{d}{\pm 2}{m}{2 \ell'}(-\beta)$; these can be straightforwardly derived using the same procedure as for the kernels $\Kk{d}{0}{m}{\ell \ell'}(-\beta)$ but with a few comments. Since $\Kk{d}{s}{m}{\ell \ell'}(-\beta)=(-1)^{\ell+\ell'} \Kk{d}{-s}{m}{\ell \ell'}(\beta)$ \citep[see Eq.~(8) of][]{Dai2014} we only need to compute the elements for $s=+2$. In addition, since $\Kk{d}{-s}{-m}{\ell \ell'}(-\beta)=\Kk{d}{s}{m}{\ell \ell'}(-\beta)$ it also follows that $\Kk{d}{s}{-m}{\ell \ell'}(-\beta)\equiv \Kk{d}{-s}{m}{\ell \ell'}(-\beta)=(-1)^{\ell+\ell'} \Kk{d}{s}{m}{\ell \ell'}(\beta)$ meaning that we only need to compute the cases for $m\geq 0$.
The required initial conditions are then
%---------------------------
\begin{align}
\Kk{d}{2}{0}{22}(-\beta)
&= \int  \frac{{_{-2}}Y{_{20}^*}(\vghp)\, {_{-2}}Y_{20}(\vgh[\vghp])}{[\gamma(1+\beta \mu')]^d} \id\vghp \equiv \frac{15}{16}\int  \frac{(1-{\mu'}^2)^2}{[\gamma(1+\beta \mu')]^{d+2}} \id\mu'
\nonumber \\
\Kk{d}{2}{0}{22}(-\beta)
&= \int  \frac{{_{-2}}Y{_{21}^*}(\vghp)\, {_{-2}}Y_{21}(\vgh[\vghp])}{[\gamma(1+\beta \mu')]^d} \id\vghp \equiv \frac{5}{8} p_{+}\int  \frac{(1-{\mu'})(1+{\mu'})^3}{[\gamma(1+\beta \mu')]^{d+2}} \id\mu'
\nonumber \\
\Kk{d}{2}{0}{22}(-\beta)
&= \int  \frac{{_{-2}}Y{_{22}^*}(\vghp)\, {_{-2}}Y_{22}(\vgh[\vghp])}{[\gamma(1+\beta \mu')]^d} \id\vghp \equiv \frac{5}{32} p^2_{+}\int  \frac{(1+{\mu'})^4}{[\gamma(1+\beta \mu')]^{d+2}} \id\mu'
\end{align}
%---------------------------
where we used $\mu[\mu']=(\mu'+\beta)/(1+\beta \mu')$,  $p=\gamma\beta$ and $p_\pm =\gamma \pm p$. The integrals are most easily solved by defining $\xi=\gamma(1+\beta \mu')=\gamma+p \mu'$ and then integrating over $\id \mu'=\id \xi/p$. We then encounter the integrals
%---------------------------
\begin{align}
\mathcal{I}_{k}(p)&=\frac{p_+^k-p_-^k}{k}=\mathcal{I}_{-k}(p)=-\mathcal{I}_{k}(-p)=\frac{\left(\frac{p_+}{p_-}\right)^{k/2}-\left(\frac{p_-}{p_+}\right)^{k/2}}{k}\equiv \frac{2\sinh\left( \frac{k}{2} \ln \left[\frac{1+p/\gamma}{1-p/\gamma}\right]\right)}{k}.
\end{align}
%---------------------------
These follow the recursion relation %---------------------------
\begin{align}
(k+1)\mathcal{I}_{k+1}(p)=2\gamma k \mathcal{I}_{k}(p)-(k-1)\mathcal{I}_{k-1}(p),
\end{align}
%---------------------------
which allows us to simply expressions. We then write
%---------------------------
\begin{align}
\Kk{d}{2}{0}{22}(-\beta)
&= \frac{15}{16 p^5}
\left[
\mathcal{I}_{1+d}(p)
-4\gamma \mathcal{I}_{d}(p)
+2(1+2\gamma^2)\mathcal{I}_{1-d}(p)
-4\gamma \mathcal{I}_{2-d}
+\mathcal{I}_{3-d}
\right]
\nonumber \\
\Kk{d}{2}{0}{22}(-\beta)
&= -\frac{5}{8 p^5}
\left[
p_- \mathcal{I}_{1+d}(p)
-(3+p^2_-) \mathcal{I}_{d}(p)
+6\gamma\mathcal{I}_{1-d}(p)
-(3+p^2_+)\mathcal{I}_{2-d}
+p_+ \mathcal{I}_{3-d}
\right]
\nonumber \\
\Kk{d}{2}{0}{22}(-\beta)
&= \frac{5}{32 p^5}
\left[
p_-^2\mathcal{I}_{1+d}(p)
-4 p_- \mathcal{I}_{d}(p)
+6 \mathcal{I}_{1-d}(p)
-4 p_+\mathcal{I}_{2-d}
+p_+^2 \mathcal{I}_{3-d}\right],
\end{align}
%---------------------------
where we used the identities $1-2 p p_+ = p_-^2$ and $1+2 p p_- = p_+^2$. These expressions allow us to generate all required kernels for our problem. We validated the correctness of our setup by explicit integration for all considered kernels. We also confirmed the expected symmetries of the expressions.

\subsection{Averaged Doppler operators up to $\mathcal{O}(p^4)$}
\label{app:Doppler_op_p4}
%---------------------------
Up to fourth order in $p$ for the intensity states the Doppler operators are then given by
%---------------------------
\bsub
\label{eq:D_explicit}
\bealf{
\Dbo{0 0}{000}{}&\approx 1-\left(3\oOnu-\oOnu^2\right)\frac{p^2}{3} 
+\left(8\oOnu+\frac{10\oOnu^2}{3}-4\oOnu^3+\frac{2\oOnu^4}{3}\right)\frac{p^4}{15} 
\nonumber\\
\frac{\Dbo{0 0}{101}{}}{3}&\approx -\left(\frac{2}{3}-\oOnu+\frac{\oOnu^2}{3}\right)\frac{p^2}{3} 
+\left(\frac{8}{3}-2\oOnu-\frac{7\oOnu^2}{3}+2\oOnu^3-\frac{\oOnu^4}{3}\right)\frac{p^4}{15} 
\nonumber\\
\frac{\Dbo{0 0}{202}{}}{5}&\approx -\left(\frac{2\oOnu}{5}-\frac{11\oOnu^2}{15}+\frac{2\oOnu^3}{5}-\frac{\oOnu^4}{15}\right)\frac{p^4}{15} 
\\
\frac{\Dbo{0 0}{020}{}}{10}&\approx \left(\frac{2\oOnu}{5}+\frac{\oOnu^2}{6}-\frac{\oOnu^3}{5}+\frac{\oOnu^4}{30}\right)\frac{p^4}{15} 
\nonumber\\
\frac{\Dbo{0 0}{121}{}}{30}&\approx \left(\frac{4}{15}+\frac{\oOnu}{5}-\frac{\oOnu^2}{15}\right)\frac{p^2}{3}
-\left(\frac{28}{15}+\frac{11\oOnu}{5}-\frac{2\oOnu^2}{15}-\frac{2\oOnu^3}{5}+\frac{\oOnu^4}{15}\right)\frac{p^4}{15} 
\nonumber\\
\frac{\Dbo{0 0}{222}{}}{50}&\approx \frac{1}{10}-\left(\frac{3}{5}+\frac{3\oOnu}{10}-\frac{\oOnu^2}{10}\right)\frac{p^2}{3}
+\left(\frac{144}{35}+\frac{17\oOnu}{5}-\frac{47\oOnu^2}{105}-\frac{16\oOnu^3}{35}+\frac{8\oOnu^4}{105}\right)\frac{p^4}{15} 
\nonumber\\
\frac{\Dbo{0 0}{323}{}}{70}&\approx \left(\frac{6}{35}+\frac{9\oOnu}{70}-\frac{3\oOnu^2}{70}\right)\frac{p^2}{3}
-\left(\frac{72}{35}+\frac{72\oOnu}{35}-\frac{3\oOnu^2}{10}-\frac{9\oOnu^3}{35}+\frac{3\oOnu^4}{70}\right)\frac{p^4}{15} 
\nonumber\\
\frac{\Dbo{0 0}{424}{}}{90}&\approx \left(\frac{8}{21}+\frac{2\oOnu}{5}-\frac{\oOnu^2}{21}-\frac{2\oOnu^3}{35}+\frac{\oOnu^4}{105}\right)\frac{p^4}{15}.
}
\esub
%---------------------------
For the operators involving one polarization state we find
%---------------------------
\bsub
\label{eq:D_explicit_s0}
\bealf{
\frac{\Dbo{\pm2 0}{222}{}}{50}&\approx \frac{1}{10}-\left(\frac{2}{5}+\frac{3\oOnu}{10}-\frac{\oOnu^2}{10}\right)\frac{p^2}{3} 
+\left(\frac{16}{7}+\frac{12\oOnu}{5}-\frac{2\oOnu^2}{7}-\frac{12\oOnu^3}{35}+\frac{2\oOnu^4}{35}\right)\frac{p^4}{15} 
\nonumber\\
\frac{\Dbo{\pm2 0}{323}{}}{70}&\approx \frac{1}{\sqrt{5}}\left(\frac{2}{7}+\frac{3\oOnu}{14}-\frac{\oOnu^2}{14}\right)\frac{p^2}{3}
-\frac{1}{\sqrt{5}}\left(\frac{20}{7}+3\oOnu-\frac{5\oOnu^2}{14}-\frac{3\oOnu^3}{7}-\frac{\oOnu^4}{14}\right)\frac{p^4}{15} 
\nonumber\\
\frac{\Dbo{\pm2 0}{424}{}}{90}&\approx \frac{1}{\sqrt{15}}\left(\frac{20}{21}+\oOnu-\frac{5\oOnu^2}{42}-\frac{\oOnu^3}{7}+\frac{\oOnu^4}{42}\right)\frac{p^4}{15} 
\nonumber\\
\Dbo{0, \pm2}{\ell 2 \ell}{}
&\equiv \Dbo{\pm2 0}{\ell 2 \ell}{}.
}
\esub
%---------------------------
We were unable to explicitly prove the last identity but we confirmed it to be correct even for higher order terms. 

Finally, for the operators connecting polarization states we have 
%---------------------------
\bsub
\label{eq:D_explicit_psps}
\bealf{
\frac{\Dbo{\pm2, \pm2}{222}{}}{50}&\approx \frac{1}{10}-\left(\frac{1}{3}+\frac{\oOnu}{10}-\frac{\oOnu^2}{30}\right)\frac{p^2}{3} 
+\left(\frac{12}{7}+\frac{4\oOnu}{5}-\frac{19\oOnu^2}{105}-\frac{2\oOnu^3}{35}+\frac{\oOnu^4}{105}\right)\frac{p^4}{15} 
\nonumber\\
\frac{\Dbo{\pm2, \pm2}{323}{}}{70}&\approx \left(\frac{2}{21}+\frac{\oOnu}{14}-\frac{\oOnu^2}{42}\right)\frac{p^2}{3}
-\left(\frac{6}{7}+\frac{11\oOnu}{14}-\frac{13\oOnu^2}{84}-\frac{\oOnu^3}{14}
+\frac{\oOnu^4}{84}\right)\frac{p^4}{15} 
\nonumber\\
\frac{\Dbo{\pm2, \pm2}{424}{}}{90}&\approx \left(\frac{10}{63}+\frac{\oOnu}{6}-\frac{5\oOnu^2}{252}-\frac{\oOnu^3}{42}+\frac{\oOnu^4}{252}\right)\frac{p^4}{15} 
\\[2mm]
\frac{\Dbo{\pm2, \mp2}{222}{}}{50}&\approx \frac{\Dbo{\pm2, \pm2}{222}{}}{50}+\left(\frac{4}{15}-\frac{2\oOnu}{5}+\frac{2\oOnu^2}{15}\right)\frac{p^2}{3} 
-\left(\frac{4}{3}-\frac{6\oOnu}{5}-\frac{4\oOnu^2}{5}+\frac{4\oOnu^3}{5}-\frac{2\oOnu^4}{15}\right)\frac{p^4}{15} 
\nonumber\\
\frac{\Dbo{\pm2, \mp2}{323}{}}{70}&\approx \frac{\Dbo{\pm2, \pm2}{323}{}}{70}
+\left(\frac{4}{21}-\frac{\oOnu}{7}-\frac{\oOnu^2}{6}+\frac{\oOnu^3}{7}-\frac{\oOnu^4}{42}\right)\frac{p^4}{15} 
\nonumber\\
\frac{\Dbo{\pm2, \mp2}{424}{}}{90}&\approx \frac{\Dbo{\pm2, \pm2}{424}{}}{90}.
}
\esub
%---------------------------
All other operators being higher order in $p$. This aspect can be understood by realizing that the aberration kernel only couples multipoles at order $p^{|\ell-\ell'|}$ \citep{Chluba2011ab, Dai2014}. The obtained operators will allow us to compute results up to second order in the temperature. We will also consider more orders but without explicitly showing the intermediate steps.

\section{Gaunt integrals}
\label{app:Gaunt}
%--------------------------------------
Integrals over products of three spin-weighted spherical harmonics can be nicely written in terms of the Gaunt coefficients
%-----------
\bealf{
\label{eq:Gaunt}
{}_{s,s_1,s_2}\mathcal{G}^{\ell,\ell_1,\ell_2}_{m,m_1,m_2}
&=
\int \id \hat{\vek{r}}\,
{}_{s}Y_{\ell m}(\hat{\vek{r}})
\,{}_{s_1}Y_{\ell_1 m_1}(\hat{\vek{r}})
\,{}_{s_2}Y_{\ell_2 m_2}(\hat{\vek{r}})
\nonumber\\
&
=\sqrt{
\frac{(2\ell+1)(2\ell_1+1)(2\ell_2+1)}{4\pi}
}
\,\Bigg(
\begin{array}{ccc}
\ell  & 
\ell_1 & \ell_2\\
-s & -s_1 & -s_2
\end{array}
\Bigg)
\,\Bigg(
\begin{array}{ccc}
\ell  & \ell_1 & \ell_2\\
m & m_1 & m_2
\end{array}
\Bigg),
}
%-----------
where we used the Wigner-$3J$ symbols. These enforce $s+s_1+s_2=m+m_1+m_2=0$. 
%, which means that $\mathcal{G}^{\ell,\ell_1,\ell_2}_{m,m_1,-m_1}$ means $m=0$. Similarly, $(-1)^m \mathcal{G}^{\ell,\ell_1,\ell_2}_{-m,m_1,m_2}$ appears when the first spherical harmonic is conjugated, as often relevant to projection integrals. We also have $[\mathcal{G}^{\ell,\ell_1,\ell_2}_{m,m_1,m_2}]^*=\mathcal{G}^{\ell,\ell_1,\ell_2}_{-m,-m_1,-m_2}$.

\end{appendix}

\end{document}

%% file: Befehle.tex
\newcommand{\expf}[1]{{{\rm e}^{#1}}}

\newcommand{\nbb}{{n^{\rm pl}}}

\newcommand{\vgh}{{\hat{\boldsymbol\gamma}}}
\newcommand{\vghp}{{\hat{\boldsymbol\gamma}'}}

\newcommand{\vb}{{\boldsymbol{\beta}}}
\newcommand{\vbh}{{\boldsymbol{\hat{\beta}}}}

\newcommand{\gammac}{{\gamma_{\rm p}}}

\newcommand{\id}{{\,\rm d}}

\newcommand{\beq}{\begin{equation}}   %

\newcommand{\eeq}{\end{equation}}   %

\newcommand{\beqa}{\begin{eqnarray}}   %

\newcommand{\eeqa}{\end{eqnarray}}   %

\newcommand{\bealf}[1]{\begin{align} #1 \end{align}}

\newcommand{\beal}{\begin{align}}
\newcommand{\enal}{\end{align}}

\newcommand{\bspl}{\begin{split}}

\newcommand{\espl}{\end{split}}

\newcommand{\bsub}{\begin{subequations}}

\newcommand{\esub}{\end{subequations}}

\newcommand{\bmulti}{\begin{multline}}   %

\newcommand{\beqm}{\begin{mathletters}}   %

\newcommand{\eeqm}{\end{mathletters}}   %

\newcommand{\kB}{k_{\rm B}}

\newcommand{\me}{m_{\rm e}}

\newcommand{\Ne}{N_{\rm e}}

\newcommand{\Te}{T_{\rm e}}

\newcommand{\Tg}{T_{\gamma}}

\newcommand{\The}{\theta_{\rm e}}

\newcommand{\vek} [1]{\mbox{\boldmath${#1}$\unboldmath}}